\documentclass[aps,prx,twocolumn,floats,showpacs,superscriptaddress,nofootinbib]{revtex4-1}

\usepackage{feynmp}
\usepackage{graphicx,epsfig}
\usepackage{times}
\usepackage{graphics,dcolumn,bm,float}
\usepackage{amssymb,amsmath,rotate,color}
\usepackage[title,titletoc,toc]{appendix}
\usepackage{graphicx}
\usepackage{wrapfig}

\usepackage{lipsum}
\usepackage{mwe}

\usepackage[pagebackref=false,colorlinks,linkcolor=blue,citecolor=blue,urlcolor=magenta]{hyperref}

\usepackage[mathlines]{lineno}
\usepackage{hyperref}
\usepackage{breqn}
\usepackage{bbold}
\usepackage{pgfplots}
\usepackage{float}
\usepackage{tkz-euclide}
\usepackage{braket}
\usepackage{physics}
\usepackage[caption=false]{subfig}
\usepackage[export]{adjustbox}
\usepackage{tikz}
\usetikzlibrary{through,calc}
\usetikzlibrary{positioning}

\newcommand{\hlt}[1]{\textcolor{red}{#1}}
\newcommand{\be}{\begin{equation}}
\newcommand{\ee}{\end{equation}}
\newcommand{\ep}{\epsilon}

\newcommand{\bearr}{\begin{eqnarray}}
\newcommand{\eearr}{\end{eqnarray}}
\newcommand{\nn}{\nonumber}

\newcommand{\bsq}{{\boldsymbol{q}}}
\newcommand{\bsk}{{\boldsymbol{k}}}
\newcommand{\pr}{\partial}
\newcommand{\bs}{\boldsymbol}

\begin{document}
\preprint{}
\title{Undamped transverse electric mode in undoped two-dimensional tilted Dirac cone materials}
\author{Z. Jalali-Mola}
\email{jalali@physics.sharif.edu}
\affiliation{
Department of Physics$,$ Sharif University of  Technology$,$ Tehran 11155-9161$,$ Iran
}

\author{S.A. Jafari}
\email{jafari@physics.sharif.edu}
\affiliation{
Department of Physics$,$ Sharif University of  Technology$,$ Tehran 11155-9161$,$ Iran
}
\affiliation{Center of excellence for Complex Systems and Condensed Matter (CSCM)$,$
Sharif University of Technology$,$Tehran 1458889694$,$ Iran}

\date{\today}

\begin{abstract}
Transverse electric (TE) modes can not propagate through the conducting solids. This is because 
the continuum of particle-hole excitations of conductors contaminates with the TE mode and dampes it out.
But in solids hosting tilted Dirac cone (TDC) that admit a description in terms of a modified Minkowski spacetime,
the new spacetime structure remedies this issue and therefore a tilted Dirac cone material (TDM) 
supports the propagation of an undamped TE mode which is sustained by density fluctuations. 
The resulting TE mode propagates at fermionic velocities which strongly confines the mode to the surface of the two-dimensional (2D) TDM. 
\end{abstract}



\maketitle
\narrowtext

{\em Introduction --}
Electromagnetic waves in the vacuum can propagate in both TE and transverse magnetic (TM) modes~\cite{JacksonBook}. 
When they hit a conductor that hosts free charge carriers, the TM mode organizes itself into fluctuations of the density known as plasmons~\cite{PinesBook,vignale}. 
But the TE mode in conducting media encounters a severe obstacle: The contamination with the particle-hole continuum (PHC) of the conductor, immediately cause
the Landau damping of the TE mode~\cite{PinesBook}. In three dimensional (3D) electron liquids, within the Landau Fermi liquid approach, the theory
requires a Landau parameter $F_1^s>6$, in order to support undamped TE mode outside the PHC~\cite{PinesBook}. But such a large
value of residual Coulomb interaction is not easy to realize in 3D electron systems. 

Dirac materials are the new class of conductors where instead of one band, one deals with two bands that touch
linearly~\cite{WehlingReview,Armitage2018,CastroNeto}. Interband transitions in these systems contribute to the polarization response of the system.
In doped graphene, it has been predicted~\cite{Mikhailov2007} and experimentally observed~\cite{Menabde2016} that the unique void below the interband part of the PHC
allows the TE mode of the vacuum in a narrow frequency range to propagate through the graphene~\cite{Mikhailov2007,SaharStoner,Kotov2013,Bludov2013}. 
The dispersion relation is quite close to the light dispersion in vacuum that makes Dirac matter potentially useful for the development of a broadband 
TE-pass polarizer~\cite{jablan2009,Angilella2011,Bao2011}. Such TE modes are also expected in bilayer graphene~\cite{Jablan2011} 
in a more pronounced way~\cite{Gomez2012,Jablan2011,Stauber2012}.
In undoped 3D Weyl semimetals, an additional vector $\bs b$ enters the theory
that parametrizes the axion field~\cite{ZhangPhysToday}. This vector parameter gaps out one of the TE modes of the vacuum, but still does not cure  
the Landau damping~\cite{Alberto2016}. Therefore an {\em undamped} branch of TE mode in conducting media remains elusive. The purpose of the present letter is to 
show that, in 2D TDMs, the tilt parameter saves the TE mode from Landau damping by pulling it out of the PHC. 

When the Dirac theory comes to mundane sub-eV energy scales of the band structure of solids, 
it can be modified in many interesting ways. One fascinating deformation of the Dirac theory is to tilt the Dirac cone.
This can be achieved in a variety of systems. The first realization of such tilting was in 
\hlt{a} $\alpha$-(BEDT-TTF)$_2$I$_3$ organic \hlt{salts} under high pressure~\cite{Kajita1992,Suzumura2006,Suzumura2014}.
A pure 2D allotrope of boron with $8pmmn$ symmetry group is also predicted to be stable and possesses TDC
in its dispersion relation~\cite{Zhou2014,Zabolotskiy2016,Lopez2016,Feng2017}.
Such a TDC can also be found in quantum wells of LaAlO$_3$/LaNiO$_3$/LaAlO$_3$. Interestingly, varying the
number of LaNiO$_3$ monolayers can shift the node of the TDM~\cite{Tsymbal2018}. Smaller tilts can also be induced in 
mechanically deformed graphene~\cite{Goerbig2008} and hydrogenated graphene~\cite{Lu2016}. 
Tilted Dirac/Weyl fermions can also be realized in 3D systems ranging from transition metal dichalcogenide PtTe$_2$~\cite{Yan2017} and PdTe$_2$~\cite{Noh}, 
WTe$_2$~\cite{Soluyanov2015} to MoTe$_2$~\cite{Zhijun-2016}.
The tilt of the Dirac cone will modify many physical properties: Apart from strong anisotropic conductivity~\cite{Adagideli2019} which is 
immediately expected for a TDM, at a fundamental level, the tilt modifies the chiral anomaly~\cite{Zhang2018}. 
At the interface with vacuum, the tilt can give rise to soft surface plasmon polaritons~\cite{SaharSSP}, and
in 2D TDM, tilt induces a  kink in the dispersion~\cite{SaharTilt1,SaharTilt2}. It can also give rise to anomalous magneto-plasmon 
and Hall effects~\cite{Goerbig2014,Zyuzin2016}. 
Moreover, interplay between tilt and many body interaction in doped TDMs, enhances the tilt of Dirac cones~\cite{Zhao2019,Lee2018}.
Tilt also strongly enhances superconductivity and Josephson effect~\cite{FaraeiPAR,FaraeiCharged}. 

On top of all the exotic properties of TDMs, one key observation is that they break the emergent Lorentz symmetry~\cite{Yan2017} of Dirac cones,
and give rise to a new spacetime structure~\cite{JafariEB}. This covariance is imprinted in the polarization tensor that determines the electromagnetic
response of the system~\cite{SaharCovariance}. Therefore at a very fundamental level, the structure of spacetime in TDMs is different from 
what we are used to in traditional solid state systems~\cite{Volovik2016black,Volovik2017lifshitz,Volovik2018,SaharSSP,TohidBorophene,Hashimoto2019,Wilczek2020}.
In this work, we will analytically show that the tilt will remedy this fundamental limitation of the conducting states by sustaining
a genuine {\em undamped} TE mode that propagates at the velocity scale of electrons ($v_F$) rather than the light velocity. We also establish
that the rescue from damping has intimate connection to the new spacetime structure in TDMs. 
For concreteness, we will focus on 2D case, but the covariance of our formulation makes the extension to 3D case straightforward.
Apart from the intrinsic appeal of 2D case, our interest in 2D TDMs is motivated by our recent proposal on
the {\em tunability} of the tilt and hence the geometry of the spacetime in certain TDMs~\cite{TohidBorophene}.



\textit{Electromagnetic modes at interface:}
To be concerete, we consider a 2D matter at $z=0$ surrounded by two media with equal dielectric constant $\ep$. Imposing the 
standard boundary conditions on electric/magnetic filed components at the interface of two media gives the following traditional expressions for
the dispersion relation for the dispersion of TM and TE modes,
~\cite{Nakayama1974,Mikhailov2007,SaharStoner,Maier-Book},
\bearr
&&1+\frac{2\pi i \gamma}{\omega q^2}  \Gamma(\bsq,\omega)=0~~~~~~{\rm (TM mode)},
 \label{TM-dis.eqn}\\
&&1-\frac{2\pi i \omega}{c^2\gamma q^2} \Gamma'(\bsq,\omega)=0~~~~{\rm (TE mode)},\label{TE-dis.eqn}
\eearr
where $c$ is the light velocity, $\omega$ is the frequency at which the wave is propagating, $\bsq$ is the two dimensional
wave vector in the $xy$-plane (where the 2D matter lies) and the evanescent decay along the $z$ direction of electromagnetic field in each medium 
is encoded into $\gamma^2=q^2-\omega^2/(c^2\ep)$.
The functions $\Gamma(\bsq,\omega)$ and $\Gamma'(\bsq,\omega)$ relate to components of the conductivity tensor
\be
\bs\sigma(\bsq ,\omega)=
\begin{bmatrix}
\sigma^{xx}	(\bsq ,\omega)&	 \sigma^{xy}(\bsq ,\omega)\\
  	\sigma^{yx}(\bsq ,\omega)	&  \sigma^{yy}(\bsq ,\omega)
\end{bmatrix},
\ee
through
\bearr
&&\Gamma(\bsq,\omega)=\bra{\bsq}\bs\sigma(\bsq,\omega)\ket{\bsq}=q_x^2\sigma^{xx}(\bsq,\omega)+q_y^2\sigma^{yy}(\bsq,\omega)+\nn\\&& q_x q_y \sigma^{xy}(\bsq,\omega)+q_x q_y \sigma^{yx}(\bsq,\omega),
 \label{gamma.eqn}\\
&&\Gamma'(\bsq,\omega)=\bra{\bsq'}\bs\sigma(\bsq,\omega)\ket{\bsq'}=
q_x^2 \sigma^{yy}(\bsq ,\omega)+ q_y^2 \sigma^{xx}(\bsq ,\omega)\nn\\&&
-q_x q_y \sigma^{xy}(\bsq ,\omega) -q_x q_y \sigma^{yx}(\bsq ,\omega).
\label{gamma'.eqn}
\eearr
Here, $\ket{\bsq}^\intercal=(q_x,q_y)$ and $\ket{\bsq'}^\intercal=(-q_y,q_x)=\bs R(\pi/2) \ket{\bsq}$ 
is related by a rotation matrix $\bs R$ (along the $z$ axis) through an angle $\pi/2$ to the vector  $\ket{\bsq}^\intercal$. 
The conductivity tensor $\bs\sigma$ is the part which includes the properties of 2D matter. 
The detail of derivation is given in the supplementary material (SM). 

So far the discussion is generic. Later we will specialize \hlt{in} the case of TDMs. 
Eq.~\eqref{TM-dis.eqn} results from imposing the boundary condition on $\bs H$ field in the $xy$ plane 
(perpendicular to the propagation direction $\bsq$, hence the name TM).
In the non-retarded limit, $c\to \infty$, the TM mode reduces to the well-known result known as random phase approximation (RPA),
\be
\ep^{RPA}(\bsq,\omega)=1-v(q) \chi (\bsq,\omega)=0,
\ee
where $v(q)=2\pi e^2/q$ and $\chi (\bsq,\omega)$ is density-density correlation function and is related to the
conductivity tensor $\sigma^{ij}$ via the continuity equation~\cite{vignale}.
The solution of the above TM-mode equation represents the collective excitations of charge density  known as plasmon. 
These modes admit a straightforward intuition in the context of  the classical Drude model and Hydrodynamic treatment 
as collective charge oscillations~\cite{Fetter1973,Lucas2018}. 

 The second dispersion given by Eq.~\eqref{TE-dis.eqn}, is the solution of electromagnetic field equations for the situation where the $\bs E$ field
 lies in the plane of 2D matter. 
 Such modes despite being very popular in waveguides~\cite{Okamoto-Book},
remain elusive in electron liquids. Within the Fermi liquid theory, a very strong Landau Fermi liquid parameter $F^s_{1}>6$ is required to sustain the undamped TE mode
in 3D electron liquids. Otherwise, the TE modes in electron systems acquire Landau damping~\cite{PinesBook}. 
In doped graphene, inter-band particle-hole excitations facilitate the propagation of such mode, but the propagation velocity is
on the scale of the light velocity $c$~\cite{Mikhailov2007} and its frequency is severely limited to $1.66\lesssim\omega/\mu<2$ where the bounds are set
by the chemical potential (doping level) $\mu$ which is typically a fraction of electron-volt. Such a high velocity leads to 
poor confinement~\cite{Nakayama1974,Mikhailov2007,Bludov2013,Kotov2013,SaharStoner,Jablan2011} of the graphene-based TE modes. 
The major finding of the present paper is that the new spacetime structure of TDMs~\cite{JafariEB,SaharCovariance,SaharSSP,Feng2018,TohidBorophene,Hashimoto2019,Wilczek2020,kang2019optical,HawkingTemperatureSun} provides a genuine opportunity for the formation of TE modes that are not
limited by any doping scale and are furthermore sustained with matter currents as their velocities are on the scale of $v_F$. 
Let us now see how does Eq.~\eqref{TE-dis.eqn} admit a solution in TDMs. 

\textit{TE mode equation in TDMs}: The low energy excitations of tilted Dirac semimetals around one Dirac point in 2D are described by the following general Hamiltonian,
\begin{equation}
   H(k)=\hbar \begin{pmatrix}  v_{x,t}k_x + v_{y,t}k_y &  v_x k_x-i v_y k_y\\    v_x k_x+i v_y k_y &  v_{x,t}k_x + v_{y,t}k_y \end{pmatrix}.
     \label{Hamiltonian.eqn}
 \end{equation}
Here, $v_x$ and $v_y$ are the anisotropic Fermi velocity along $x$ and $y$  direction, respectively. Tilt of Dirac cone in each direction is defined by $v_{x,t}$ and $v_{y,t}$. 
The time reversal partner of above Dirac cone (valley), has opposite tilt direction ($-v_{x,t}$ and $-v_{y,t}$). In the case of zero tilt $v_{x,t}=v_{y,t}=0$, 
and assuming the isotropy condition $v_x=v_y=v_F$,
the above Hamiltonian reduces to the effective theory of graphene. 
Reparameterizing the tilt $\bs\zeta=(\zeta_x,\zeta_y)$ as 
$v_{x,t}=\zeta_x v_F$ and $v_{y,t}=\zeta_y v_F$, one arrives at,
\begin{equation}
   H(k)=\hbar v_F \begin{pmatrix}  \zeta_{x}k_x + \zeta_{y}k_y &   k_x-i  k_y\\     k_x+i k_y &  \zeta_{x}k_x + \zeta_{y}k_y \end{pmatrix}.
     \label{Hamiltonian1.eqn}
 \end{equation}
The energy eigenvalues $E_s=\hbar v_F k [1+s \zeta \cos(\theta_t-\theta_k)]$ describe the conduction $s=+$ and valence band $s=-$. The $\theta_k$ 
and $\theta_t$ are polar angles of the momentum and tilt vectors $\bsk,\bs\zeta$ with respect to the $x$ axis and $k=|\bsk|$ and $\zeta=|\bs\zeta|$. 
As can be seen, in the presence of the tilt, $\bs \zeta$, the energy spectrum still remains linear in wave vector $k$, but the velocity depends on the
direction. More fundamentally, it has been shown that the $\bs\zeta$ can be encoded into a suitable spacetime
metric~\cite{Volovik2016black,Volovik2017lifshitz,Volovik2018,SaharCovariance,SaharSSP,TohidBorophene,Hashimoto2019,Wilczek2020}.

What we need for the rest of the calculation is the conductivity tensor for TDC. 
It turns out that the new spacetime structure allows to express the polarization tensor in a covariant form
obeying the Ward identity~\cite{SaharCovariance}:
 \bearr
  && \Pi^{\mu \nu}= -[(q^2-\Omega^2) g^{\mu \nu}-q^{\mu} q^{\nu}]  \pi({q}^2),
  \label{tilted-cov.eqn}
\eearr
where the scalar part is $\pi({q})=-\frac{g_s v_F}{16 \hbar \sqrt{ q^2-\Omega^2}}$ and 
the non-Minkowski metric $g^{\mu\nu}$ is given by,
 \be
 g^{\mu \nu}=
 \begin{pmatrix}
 -1 &- \zeta_x & -\zeta_y \\
  -\zeta_x & 1-\zeta_x^2 & -\zeta_x \zeta_y \\
   -\zeta_y & -\zeta_x \zeta_y &  1-\zeta_y^2
 \end{pmatrix}.
 \label{metric.eqn}
 \ee
 and (2+1)-dimensional energy-momentum vectors are $q_\mu=(-\omega/v_F,\bsq)$, ${q}^\mu= g^{\mu \nu} q_{\nu}$, $q=|\bsq|$, $\Omega=\omega/v_F-\bsq.\bs\zeta$ and $g_s$ 
is the spin degeneracy. 
It is useful to explicitely represent the components as,
 \bearr
 \label{xx-yy-xy-undoped.eqn}
 &&\Pi^{00}(\boldsymbol{q},\Omega)=
   \{q^2 \} \pi(q),\\
   &&\Pi^{i0}(\boldsymbol{q},\Omega)=\tilde\Pi^{0i}(\boldsymbol{q},\Omega)=
     \{q_i \Omega+q^2\zeta_i  \}   \pi(q),
 \nn\\
    &&\Pi^{ij}(\boldsymbol{q},\Omega)=\nn\\
     &&\{(\Omega^2-q^2)\delta_{ij}+q_i q_j+q^2\zeta_i \zeta_j+\Omega q_i \zeta_j+\Omega q_j\zeta_i
     \}  \pi(q).\nn
  \eearr    
The spatial components $\Pi^{ij}$ of the above polarization is related to the conductivity tensor $\sigma^{ij}$
required in Eqns.~\eqref{gamma.eqn} and~\eqref{gamma'.eqn} by the standard relation $-i\omega \sigma^{ij}=e^2\Pi^{ij}$~\cite{wen,Solyom2010-vol3}. 
Therefore the $\Gamma$ and $\Gamma'$ functions needed in TM and TE dispersions relations~\eqref{TM-dis.eqn} and~\eqref{TE-dis.eqn} 
will become,
 \bearr
&&- i \omega \Gamma(\bsq,\omega)=e^2\omega^2 \Pi^{00}(\bsq,\Omega), \label{TMpi.eqn}\\
&&- i \omega \Gamma'(\bsq,\omega)=e^2 q^2 (\Omega^2-\bsq^2+|\bsq\times \bs\zeta|^2)\pi(q). \label{TEpi.eqn}
 \eearr   
As a consistency check, note that in the long wavelength limit $q\to 0$, Eqs.~\eqref{TMpi.eqn} and~\eqref{TEpi.eqn} reduce to corresponding
expressions derived by Mikhailov and Zeigler~\cite{Mikhailov2007} for a single layer graphene.  
In the following, let us use put the above statement in a covariant form that emphasizes the peculiar role of 
the spacetime given by metric~\eqref{metric.eqn}.  
Using the longitudinal projection operator ${P_{||}}_i^a=\hat{q}^a \hat{q}_i$ we can project the longitudinal part of the
polarization tensor, ${\Pi_{||}}^{ab}={P_{||}}^a_i\Pi^{ij} {P_{||}}^b_j$ to obtain,
\be
   \Pi^{ab}_{||}=\hat{q}^a \hat{q}^b \omega^2  \pi(q),
   \label{Long.eqn}
\ee
where as usual Latine indices stand for space indices. 
Similarly, the transverse part of the $\Pi^{ij}$ is
\be
   \Pi_{\perp}^{ab}={P_{\perp}}^a_i\Pi^{ij}{P_{\perp}}^b_{j},~~~{P_\perp}^{a}_i=\delta^a_i-\hat q^a \hat q_i
\ee
which for Eq.~\eqref{xx-yy-xy-undoped.eqn} becomes,
\bearr
   \Pi^{ab}_\perp / \pi(q)&&= (\delta^{ab}-\hat{q}^a \hat{q}^b) (\Omega^2-q^2)\\
   &&+q^2\zeta^a\zeta^b+(\bs\zeta.\hat{\bsq})^2 q^a q^b - (\bs\zeta.\bsq) \zeta^a q^b -(\bs\zeta.\bsq)\zeta^b q^a,\nn\\
   &&= (\delta^{ab}-\hat{q}^a \hat{q}^b) (\Omega^2-q^2)+(\delta^{ab}-\hat{q}^a \hat{q}^b) |\bsq\times\bs\zeta|^2.\nn
\eearr
This can be compactly rewritten as,
\be
   \Pi^{ab}_\perp=(\delta^{ab}-\hat{q}^a \hat{q}^b) ((\Omega^2-q^2)+|\bsq\times\bs\zeta|^2) \pi(q).
   \label{Trans.eqn}
\ee
This establishes that $\Gamma$ and $\Gamma'$ determining the dispersion of TM and TE modes in equations~\eqref{TMpi.eqn} and~\eqref{TEpi.eqn}
are actually proportional to longitudinal ($\Pi_{||}$) and transverse ($\Pi_{\perp}$) parts of the polarization tensor. 

The dispersion of TM modes in Eq.~\eqref{TMpi.eqn} is given by density-density correlations, which 
according to Eq.~\eqref{Long.eqn} is determined by the longitudinal part of the polarization. These are
the well-known plasmons modes. 
The only difference with graphene is that instead of $\omega$, we have $\Omega=\omega/v_F-\bs \zeta.\bs q$. 
It is well-known that in the undoped case, this equation does not admit a solution in the singlet channel~\cite{Wunsch2006,Hwang2007,Novoselov2012},
although a solution can be generated by including the Stoner fluctuations corresponding to ladder diagrams~\cite{SaharStoner,Mishchenko},
or by including thermal fluctuations to generate thermoplasmons~\cite{Vafek2006,Sarma2013}. 
For the same reason, the TM mode in undoped tilted Dirac system does not have a solution. 
Doping the Dirac cone changes this picture. 
Similar to upright Dirac cone, the TM mode (being density oscillation) can be sustained only in presence of a non-zero density, i.e. it requires
to dope the TDM, e.g. by gate voltage~\cite{Novoselov2004gate,Wang2008gated-transistor,Chen2012optical} or 
chemical functionalization~\cite{Katsnelson2008defect,Boukhvalov2009,Kuila2012chemical-functionalization}. The essential difference of the TM mode in TDC systems, 
with respect to the upright Dirac cone systems, is that in the presence of the tilt, a kink in the dispersion of plasmons emerges~\cite{SaharTilt1,SaharTilt2}.
This kink appears at the boundary of intra-band PHC. Furthermore, an additional overdamped plasmon branch appears
which due contamination with intra-band part of the PHC is heavily Landau-damped~\cite{SaharTilt1,SaharTilt2}. 
So the TM mode in upright and TDCs  are somehow similar. 

The quantity $\Gamma'$ determining the TE mode equation~\eqref{TEpi.eqn} or equivalently, Eq.~\eqref{Trans.eqn}
is heavily affected by the presence of the tilt term. The tilt appears not only in $\Omega$, but also as an explicit 
$\bs q \times \bs \zeta$ term that stems from the covariant structure $\Pi^{\mu\nu}$ of Eq.~\eqref{tilted-cov.eqn} in
the spacetime~\eqref{metric.eqn}. In the following, we will show that the presence of $\bs q\times \bs \zeta$ will enable the undoped TDMs
to sustain an undamped TE mode.

\textit{Propagation of TE mode in two dimensional TDMs:}
First, let us argue what is special about the {\em undoped} TDC systems:
Mathematically,  Eq.~\eqref{TE-dis.eqn} may admit a propagating TE mode solution when the imaginary part of $\Gamma'(\bsq,\omega)$ is negative. 
Let us consider the doped case first. In 2D Dirac systems possessing upright Dirac cone (such as graphene), upon doping away from the Dirac node, a triangular shaped
window below the inter-band and above the intra-band continuum of particle-hole excitations appears which provides a chance for the formation of TE modes~\cite{Mikhailov2007,SaharStoner}. At $q=0$ the formation of a solution for the TE mode can be understood as follows: Around the entire Fermi surface, at the excitation energy of $2\mu$ there will
be plenty of particle-hole excitations that generate a negative and logarithmically singular imaginary part for the $\Gamma'$~\cite{Mikhailov2007}. 
In the case of TDC, at $q=0$, depending on the angle at which initial $\bs k$ state from valence band is picked, the excitation energy 
will deviate from $2\mu$. This dispersion around $2\mu$ will destroy the logarithmic singularity.
Therefore the TE mode of graphene will evolve in TDM in two respects: (i) its velocity will be closer to $c$ and
(ii) its frequency range will be even more limited. So Doping the TDM does not give any TM mode better than the one in graphene~\footnote{We have checked this numerically using our analytic solution for the tensor $\Pi^{\mu\nu}$ in doped TDC systems.}. 

Therefore we are lead to consider a 2D TDC system at charge neutrality point. 
In this case, $\Gamma'$ will be given by Eq.~\eqref{TEpi.eqn} which gives the TE mode equation,
\bearr
1+\frac{2\pi e^2}{c^2\gamma}(\Omega^2-\bsq^2+|\bsq\times \bs\zeta|^2) \pi(q)\Theta(q-\Omega)=0.
\label{FTE-dis.eqn}
\eearr
Here, we added a step function to the definition of $\Gamma'$ to pick a region in which the imaginary part of $\Pi^{ij}$
in Eq.~\eqref{xx-yy-xy-undoped.eqn} (and hence real part of $\Gamma'$) is zero.
To understand the structure of Eq.~\eqref{FTE-dis.eqn}, let us first consider the case of upright Dirac cone with $\zeta\to 0$.
In this case we have the following equation for the TE modes in undoped graphene:
 \bearr
 &&1+\frac{2\pi e^2}{c^2\gamma}(\omega^2/v_F^2-q^2)\pi(q)\Theta(v_Fq-\omega)=0.
 \label{FTE-dis-Dirac.eqn}
 \eearr
The second term in the above equation is non-zero only when $v_Fq>\omega$. In this regime $\omega^2-v_F^2q^2$ is negative. But since $\pi(q)$ itself will also
be negative, the above equation admits no solution. That is why Mikhailov {\em et. al.} have considered the doped graphene
to host TE modes~\cite{Mikhailov2007}. Although the part of transverse polarization~\eqref{Trans.eqn} that does not involve $\bs q\times \bs \zeta$,
does indeed produces a transverse electric field polarization, but this mode does not survive the contamination with the (inter-band) PHC of the material~\footnote{
If there was no material, there would be no contamination, and therefore this part of $\Pi_\perp$ would be enough to let the TE mode propagate in vacuum.}
and thereby acquires Landau damping which prevents the propagation of the TE mode. 
We are now prepared to understand how does the $\bs q\times \bs \zeta$ term arising from the structure of the spacetime 
comes to rescue from the contamination, whereby giving rise to a genuine TE mode sustained by the fluctuations of charge fluctuations of the material. 
 
Substituting the value of $\pi(q)$ in Eq.~\eqref{FTE-dis.eqn} one obtains,
  \be
  c \gamma -\frac{\pi \alpha\Theta(q-\Omega)}{4\sqrt{\bsq^2-\Omega^2}} (\Omega^2-q^2+|\bsq\times \bs\zeta|^2),
  \label{TE-expand.eqn}
  \ee
where for clarity, we assume TDM is immersed in vacuum with $\ep=1$. The $\alpha=e^2/\hbar c=1/137$ is the fine structure constant 
and $c\gamma=\sqrt{(c/v_F)^2(v_Fq)^2-\omega^2}$ determines the confinement of the mode to 2D material. 

For $\Omega>q$ (inside the inter-band PHC), one trivial (and damped) solution is given by $\gamma=0$, 
or equivalently $\omega=c q$. This is the TE mode of vacuum which is damped by the presence of the gapless material. 
For $q>\Omega$ where the second term in the above equation is non-zero, there is another genuine solution.
Perturbatively expanding in powers of $\alpha$ we obtain, 
\be
\Omega= q-\frac{q}{16}( \alpha^{2}|\hat{\bsq}\times \bs\zeta|^4).
\ee
There are two important points to be noted here: (i), the principal velocity scale of this TE mode
is set by $v_F$ and not by $c$. This readily implies $\gamma\sim q$ and therefore the confinement length is set by the wavelength. 
This can be useful in opto-electronic applications~\cite{Vakil2011,Novoselov2012}. 
The fact that the present TE mode is organized by matter velocity scales, qualifies it as a collective mode in TDMs. 
(ii) Second important point is that clearly the $\bs q\times \bs \zeta$ term arising from the structure of the spacetime has brought the energy of 
this TE mode below the PHC. 
This is how the tilt can generate a genuine and confinable TE mode that is sustained by
charge fluctuations in the material. Notice that when the propagation direction is along the tilt direction, $\bs\zeta$, 
the TE mode will be degenerated with the lower border of the PHC. This effect can be utilized to optically identify the
tilt direction, as the direction $\bs\zeta$ will be the only direction along which our TE mode ceases to propagate. 

{\em Summary and discussion:}
In this paper starting from a covariant expression for the polarization of an {\em undoped} two-dimensional TDM 
in a spacetime given by metric~\eqref{metric.eqn}, we have investigated the formation of a genuine
TE mode. The TE mode equation is modified by an important term $\bs q \times \bs\zeta$ which arises from the structure of
the spacetime. This term helps the TE mode to escape from the contamination with the PHC of the TDM. 
The velocity of the resulting mode is determined by $v_F$. Therefore in contrast to the case of graphene~\cite{Mikhailov2007},
the present TE mode will be properly confined to the 2D TDM which can be useful in opto-electronic applications~\cite{Musa2017}.
Further advantage of our scenario for a matter sustained TE mode is that,
while in graphene the TE mode is very fragile and goes away by turning on a difference in the dielectric constants
of the surrounding media~\cite{Guinea2013}, our TE mode being sustained by matter density fluctuations, does not care about 
a possible difference in the dielectric constants of the surrounding media. This proves to be very
convenient, sin the present TE mode can propagate in a TDM on a substrate. 
An important feature of our TE mode at the charge neutrality point of the TDM is that 
the {\em fluctuations} of the charge density are enough to sustain the TE mode.
One does not need the {\em average} density of the (free) charge carriers to be non-zero. 


Let us now ponder about the other valley which is inevitably present in TDM and corresponds to $-\bs\zeta$.
But since the transverse polarization involves the square of $\bs q\times \bs\zeta$, the effect of the other
valley does not wash out the TE mode. The most manifest form of this effect can take place for $\bs q \perp \bs \zeta$. 
The question of possible inter-valley collective modes remains open and pertinent. 
Existence of undamped TE modes in the 3D electron liquids requires a very strong Landau parameter $F_1^s>6$~\cite{Pines1951,PinesBook,Matsumoto1980}.
Such a quantum liquid will be so strongly correlated that, it is likely to undergo a phase transition well before reaching such a large value of
interaction parameters. From this perspective, the advantage of our proposal is that in our case, no strong Coulomb interactions are required.

\bibliography{mybib}

\begin{thebibliography}{79}%
\makeatletter
\providecommand \@ifxundefined [1]{%
 \@ifx{#1\undefined}
}%
\providecommand \@ifnum [1]{%
 \ifnum #1\expandafter \@firstoftwo
 \else \expandafter \@secondoftwo
 \fi
}%
\providecommand \@ifx [1]{%
 \ifx #1\expandafter \@firstoftwo
 \else \expandafter \@secondoftwo
 \fi
}%
\providecommand \natexlab [1]{#1}%
\providecommand \enquote  [1]{``#1''}%
\providecommand \bibnamefont  [1]{#1}%
\providecommand \bibfnamefont [1]{#1}%
\providecommand \citenamefont [1]{#1}%
\providecommand \href@noop [0]{\@secondoftwo}%
\providecommand \href [0]{\begingroup \@sanitize@url \@href}%
\providecommand \@href[1]{\@@startlink{#1}\@@href}%
\providecommand \@@href[1]{\endgroup#1\@@endlink}%
\providecommand \@sanitize@url [0]{\catcode `\\12\catcode `\$12\catcode
  `\&12\catcode `\#12\catcode `\^12\catcode `\_12\catcode `\%12\relax}%
\providecommand \@@startlink[1]{}%
\providecommand \@@endlink[0]{}%
\providecommand \url  [0]{\begingroup\@sanitize@url \@url }%
\providecommand \@url [1]{\endgroup\@href {#1}{\urlprefix }}%
\providecommand \urlprefix  [0]{URL }%
\providecommand \Eprint [0]{\href }%
\providecommand \doibase [0]{http://dx.doi.org/}%
\providecommand \selectlanguage [0]{\@gobble}%
\providecommand \bibinfo  [0]{\@secondoftwo}%
\providecommand \bibfield  [0]{\@secondoftwo}%
\providecommand \translation [1]{[#1]}%
\providecommand \BibitemOpen [0]{}%
\providecommand \bibitemStop [0]{}%
\providecommand \bibitemNoStop [0]{.\EOS\space}%
\providecommand \EOS [0]{\spacefactor3000\relax}%
\providecommand \BibitemShut  [1]{\csname bibitem#1\endcsname}%
\let\auto@bib@innerbib\@empty
\bibitem [{\citenamefont {Jackson}(1999)}]{JacksonBook}%
  \BibitemOpen
  \bibfield  {author} {\bibinfo {author} {\bibfnamefont {J.~D.}\ \bibnamefont
  {Jackson}},\ }\href@noop {} {\enquote {\bibinfo {title} {Classical
  electrodynamics},}\ } (\bibinfo {year} {1999})\BibitemShut {NoStop}%
\bibitem [{\citenamefont {Pines}\ and\ \citenamefont
  {Nozières}(1966)}]{PinesBook}%
  \BibitemOpen
  \bibfield  {author} {\bibinfo {author} {\bibfnamefont {D.}~\bibnamefont
  {Pines}}\ and\ \bibinfo {author} {\bibfnamefont {P.}~\bibnamefont
  {Nozières}},\ }\href@noop {} {\emph {\bibinfo {title} {Theory of Quantum
  Liquids: Normal Fermi Liquids}}}\ (\bibinfo  {publisher} {Avalon
  Publishing},\ \bibinfo {year} {1966})\BibitemShut {NoStop}%
\bibitem [{\citenamefont {Giuliani}\ and\ \citenamefont
  {Vignale}(2005)}]{vignale}%
  \BibitemOpen
  \bibfield  {author} {\bibinfo {author} {\bibfnamefont {G.}~\bibnamefont
  {Giuliani}}\ and\ \bibinfo {author} {\bibfnamefont {G.}~\bibnamefont
  {Vignale}},\ }\href@noop {} {\emph {\bibinfo {title} {Quantum Theory of The
  Electron Liquid}}}\ (\bibinfo  {publisher} {Cambridge University Press},\
  \bibinfo {year} {2005})\BibitemShut {NoStop}%
\bibitem [{\citenamefont {Wehling}\ \emph {et~al.}(2014)\citenamefont
  {Wehling}, \citenamefont {Black-Schaffer},\ and\ \citenamefont
  {Balatsky}}]{WehlingReview}%
  \BibitemOpen
  \bibfield  {author} {\bibinfo {author} {\bibfnamefont {T.~O.}\ \bibnamefont
  {Wehling}}, \bibinfo {author} {\bibfnamefont {A.~M.}\ \bibnamefont
  {Black-Schaffer}}, \ and\ \bibinfo {author} {\bibfnamefont {A.~V.}\
  \bibnamefont {Balatsky}},\ }\href {\doibase 10.1080/00018732.2014.927109}
  {\bibfield  {journal} {\bibinfo  {journal} {Adv. Phys.}\ }\textbf {\bibinfo
  {volume} {63}},\ \bibinfo {pages} {1} (\bibinfo {year} {2014})}\BibitemShut
  {NoStop}%
\bibitem [{\citenamefont {Armitage}\ \emph {et~al.}(2018)\citenamefont
  {Armitage}, \citenamefont {Mele},\ and\ \citenamefont
  {Vishwanath}}]{Armitage2018}%
  \BibitemOpen
  \bibfield  {author} {\bibinfo {author} {\bibfnamefont {N.~P.}\ \bibnamefont
  {Armitage}}, \bibinfo {author} {\bibfnamefont {E.~J.}\ \bibnamefont {Mele}},
  \ and\ \bibinfo {author} {\bibfnamefont {A.}~\bibnamefont {Vishwanath}},\
  }\href {\doibase 10.1103/RevModPhys.90.015001} {\bibfield  {journal}
  {\bibinfo  {journal} {Rev. Mod. Phys.}\ }\textbf {\bibinfo {volume} {90}},\
  \bibinfo {pages} {015001} (\bibinfo {year} {2018})}\BibitemShut {NoStop}%
\bibitem [{\citenamefont {Castro~Neto}\ \emph {et~al.}(2009)\citenamefont
  {Castro~Neto}, \citenamefont {Guinea}, \citenamefont {Peres}, \citenamefont
  {Novoselov},\ and\ \citenamefont {Geim}}]{CastroNeto}%
  \BibitemOpen
  \bibfield  {author} {\bibinfo {author} {\bibfnamefont {A.~H.}\ \bibnamefont
  {Castro~Neto}}, \bibinfo {author} {\bibfnamefont {F.}~\bibnamefont {Guinea}},
  \bibinfo {author} {\bibfnamefont {N.~M.~R.}\ \bibnamefont {Peres}}, \bibinfo
  {author} {\bibfnamefont {K.~S.}\ \bibnamefont {Novoselov}}, \ and\ \bibinfo
  {author} {\bibfnamefont {A.~K.}\ \bibnamefont {Geim}},\ }\href {\doibase
  10.1103/RevModPhys.81.109} {\bibfield  {journal} {\bibinfo  {journal} {Rev.
  Mod. Phys.}\ }\textbf {\bibinfo {volume} {81}},\ \bibinfo {pages} {109}
  (\bibinfo {year} {2009})}\BibitemShut {NoStop}%
\bibitem [{\citenamefont {Mikhailov}\ and\ \citenamefont
  {Ziegler}(2007)}]{Mikhailov2007}%
  \BibitemOpen
  \bibfield  {author} {\bibinfo {author} {\bibfnamefont {S.~A.}\ \bibnamefont
  {Mikhailov}}\ and\ \bibinfo {author} {\bibfnamefont {K.}~\bibnamefont
  {Ziegler}},\ }\href {\doibase 10.1103/PhysRevLett.99.016803} {\bibfield
  {journal} {\bibinfo  {journal} {Phys. Rev. Lett.}\ }\textbf {\bibinfo
  {volume} {99}},\ \bibinfo {pages} {016803} (\bibinfo {year}
  {2007})}\BibitemShut {NoStop}%
\bibitem [{\citenamefont {Menabde}\ \emph {et~al.}(2016)\citenamefont
  {Menabde}, \citenamefont {Mason}, \citenamefont {Kornev}, \citenamefont
  {Lee},\ and\ \citenamefont {Park}}]{Menabde2016}%
  \BibitemOpen
  \bibfield  {author} {\bibinfo {author} {\bibfnamefont {S.~G.}\ \bibnamefont
  {Menabde}}, \bibinfo {author} {\bibfnamefont {D.~R.}\ \bibnamefont {Mason}},
  \bibinfo {author} {\bibfnamefont {E.~E.}\ \bibnamefont {Kornev}}, \bibinfo
  {author} {\bibfnamefont {C.}~\bibnamefont {Lee}}, \ and\ \bibinfo {author}
  {\bibfnamefont {N.}~\bibnamefont {Park}},\ }\href {\doibase
  https://doi.org/10.1038/srep21523} {\bibfield  {journal} {\bibinfo  {journal}
  {Scientific reports}\ }\textbf {\bibinfo {volume} {6}},\ \bibinfo {pages}
  {21523} (\bibinfo {year} {2016})}\BibitemShut {NoStop}%
\bibitem [{\citenamefont {Jalali-Mola}\ and\ \citenamefont
  {Jafari}(2019{\natexlab{a}})}]{SaharStoner}%
  \BibitemOpen
  \bibfield  {author} {\bibinfo {author} {\bibfnamefont {Z.}~\bibnamefont
  {Jalali-Mola}}\ and\ \bibinfo {author} {\bibfnamefont {S.}~\bibnamefont
  {Jafari}},\ }\href {\doibase https://doi.org/10.1016/j.jmmm.2018.09.040}
  {\bibfield  {journal} {\bibinfo  {journal} {Journal of Magnetism and Magnetic
  Materials}\ }\textbf {\bibinfo {volume} {471}},\ \bibinfo {pages} {220 }
  (\bibinfo {year} {2019}{\natexlab{a}})}\BibitemShut {NoStop}%
\bibitem [{\citenamefont {Kotov}\ \emph {et~al.}(2013)\citenamefont {Kotov},
  \citenamefont {Kol'chenko},\ and\ \citenamefont {Lozovik}}]{Kotov2013}%
  \BibitemOpen
  \bibfield  {author} {\bibinfo {author} {\bibfnamefont {O.}~\bibnamefont
  {Kotov}}, \bibinfo {author} {\bibfnamefont {M.}~\bibnamefont {Kol'chenko}}, \
  and\ \bibinfo {author} {\bibfnamefont {Y.~E.}\ \bibnamefont {Lozovik}},\
  }\href {\doibase 10.1364/OE.21.013533} {\bibfield  {journal} {\bibinfo
  {journal} {Opt. Express}\ }\textbf {\bibinfo {volume} {21}},\ \bibinfo
  {pages} {13533} (\bibinfo {year} {2013})}\BibitemShut {NoStop}%
\bibitem [{\citenamefont {BLUDOV}\ \emph {et~al.}(2013)\citenamefont {BLUDOV},
  \citenamefont {FERREIRA}, \citenamefont {PERES},\ and\ \citenamefont
  {VASILEVSKIY}}]{Bludov2013}%
  \BibitemOpen
  \bibfield  {author} {\bibinfo {author} {\bibfnamefont {Y.~V.}\ \bibnamefont
  {BLUDOV}}, \bibinfo {author} {\bibfnamefont {A.}~\bibnamefont {FERREIRA}},
  \bibinfo {author} {\bibfnamefont {N.~M.~R.}\ \bibnamefont {PERES}}, \ and\
  \bibinfo {author} {\bibfnamefont {M.~I.}\ \bibnamefont {VASILEVSKIY}},\
  }\href {\doibase 10.1142/S0217979213410014} {\bibfield  {journal} {\bibinfo
  {journal} {International Journal of Modern Physics B}\ }\textbf {\bibinfo
  {volume} {27}},\ \bibinfo {pages} {1341001} (\bibinfo {year}
  {2013})}\BibitemShut {NoStop}%
\bibitem [{\citenamefont {Jablan}\ \emph {et~al.}(2009)\citenamefont {Jablan},
  \citenamefont {Buljan},\ and\ \citenamefont {Solja\ifmmode \check{c}\else
  \v{c}\fi{}i\ifmmode~\acute{c}\else \'{c}\fi{}}}]{jablan2009}%
  \BibitemOpen
  \bibfield  {author} {\bibinfo {author} {\bibfnamefont {M.}~\bibnamefont
  {Jablan}}, \bibinfo {author} {\bibfnamefont {H.}~\bibnamefont {Buljan}}, \
  and\ \bibinfo {author} {\bibfnamefont {M.}~\bibnamefont {Solja\ifmmode
  \check{c}\else \v{c}\fi{}i\ifmmode~\acute{c}\else \'{c}\fi{}}},\ }\href
  {\doibase 10.1103/PhysRevB.80.245435} {\bibfield  {journal} {\bibinfo
  {journal} {Phys. Rev. B}\ }\textbf {\bibinfo {volume} {80}},\ \bibinfo
  {pages} {245435} (\bibinfo {year} {2009})}\BibitemShut {NoStop}%
\bibitem [{\citenamefont {Pellegrino}\ \emph {et~al.}(2011)\citenamefont
  {Pellegrino}, \citenamefont {Angilella},\ and\ \citenamefont
  {Pucci}}]{Angilella2011}%
  \BibitemOpen
  \bibfield  {author} {\bibinfo {author} {\bibfnamefont {F.~M.~D.}\
  \bibnamefont {Pellegrino}}, \bibinfo {author} {\bibfnamefont {G.~G.~N.}\
  \bibnamefont {Angilella}}, \ and\ \bibinfo {author} {\bibfnamefont
  {R.}~\bibnamefont {Pucci}},\ }\href {\doibase 10.1103/PhysRevB.84.195407}
  {\bibfield  {journal} {\bibinfo  {journal} {Phys. Rev. B}\ }\textbf {\bibinfo
  {volume} {84}},\ \bibinfo {pages} {195407} (\bibinfo {year}
  {2011})}\BibitemShut {NoStop}%
\bibitem [{\citenamefont {Bao}\ \emph {et~al.}(2011)\citenamefont {Bao},
  \citenamefont {Zhang}, \citenamefont {Wang}, \citenamefont {Ni},
  \citenamefont {Lim}, \citenamefont {Wang}, \citenamefont {Tang},\ and\
  \citenamefont {Loh}}]{Bao2011}%
  \BibitemOpen
  \bibfield  {author} {\bibinfo {author} {\bibfnamefont {Q.}~\bibnamefont
  {Bao}}, \bibinfo {author} {\bibfnamefont {H.}~\bibnamefont {Zhang}}, \bibinfo
  {author} {\bibfnamefont {B.}~\bibnamefont {Wang}}, \bibinfo {author}
  {\bibfnamefont {Z.}~\bibnamefont {Ni}}, \bibinfo {author} {\bibfnamefont
  {C.~H. Y.~X.}\ \bibnamefont {Lim}}, \bibinfo {author} {\bibfnamefont
  {Y.}~\bibnamefont {Wang}}, \bibinfo {author} {\bibfnamefont {D.~Y.}\
  \bibnamefont {Tang}}, \ and\ \bibinfo {author} {\bibfnamefont {K.~P.}\
  \bibnamefont {Loh}},\ }\href {\doibase 10.1038/nphoton.2011.102} {\bibfield
  {journal} {\bibinfo  {journal} {Nature Photonics}\ }\textbf {\bibinfo
  {volume} {5}},\ \bibinfo {pages} {411} (\bibinfo {year} {2011})}\BibitemShut
  {NoStop}%
\bibitem [{\citenamefont {Jablan}\ \emph {et~al.}(2011)\citenamefont {Jablan},
  \citenamefont {Buljan},\ and\ \citenamefont {Solja\v{c}i\'{c}}}]{Jablan2011}%
  \BibitemOpen
  \bibfield  {author} {\bibinfo {author} {\bibfnamefont {M.}~\bibnamefont
  {Jablan}}, \bibinfo {author} {\bibfnamefont {H.}~\bibnamefont {Buljan}}, \
  and\ \bibinfo {author} {\bibfnamefont {M.}~\bibnamefont {Solja\v{c}i\'{c}}},\
  }\href {\doibase 10.1364/OE.19.011236} {\bibfield  {journal} {\bibinfo
  {journal} {Opt. Express}\ }\textbf {\bibinfo {volume} {19}},\ \bibinfo
  {pages} {11236} (\bibinfo {year} {2011})}\BibitemShut {NoStop}%
\bibitem [{\citenamefont {G{\'{o}}mez-Santos}\ and\ \citenamefont
  {Stauber}(2012)}]{Gomez2012}%
  \BibitemOpen
  \bibfield  {author} {\bibinfo {author} {\bibfnamefont {G.}~\bibnamefont
  {G{\'{o}}mez-Santos}}\ and\ \bibinfo {author} {\bibfnamefont
  {T.}~\bibnamefont {Stauber}},\ }\href {\doibase 10.1209/0295-5075/99/27006}
  {\bibfield  {journal} {\bibinfo  {journal} {{EPL} (Europhysics Letters)}\
  }\textbf {\bibinfo {volume} {99}},\ \bibinfo {pages} {27006} (\bibinfo {year}
  {2012})}\BibitemShut {NoStop}%
\bibitem [{\citenamefont {Stauber}\ and\ \citenamefont
  {G\'omez-Santos}(2012)}]{Stauber2012}%
  \BibitemOpen
  \bibfield  {author} {\bibinfo {author} {\bibfnamefont {T.}~\bibnamefont
  {Stauber}}\ and\ \bibinfo {author} {\bibfnamefont {G.}~\bibnamefont
  {G\'omez-Santos}},\ }\href {\doibase 10.1103/PhysRevB.85.075410} {\bibfield
  {journal} {\bibinfo  {journal} {Phys. Rev. B}\ }\textbf {\bibinfo {volume}
  {85}},\ \bibinfo {pages} {075410} (\bibinfo {year} {2012})}\BibitemShut
  {NoStop}%
\bibitem [{\citenamefont {Qi}\ and\ \citenamefont
  {Zhang}(2010)}]{ZhangPhysToday}%
  \BibitemOpen
  \bibfield  {author} {\bibinfo {author} {\bibfnamefont {X.-L.}\ \bibnamefont
  {Qi}}\ and\ \bibinfo {author} {\bibfnamefont {S.-C.}\ \bibnamefont {Zhang}},\
  }\href {\doibase 10.1063/1.3293411} {\bibfield  {journal} {\bibinfo
  {journal} {Physics Today}\ }\textbf {\bibinfo {volume} {63}},\ \bibinfo
  {pages} {33} (\bibinfo {year} {2010})}\BibitemShut {NoStop}%
\bibitem [{\citenamefont {Ferreiros}\ and\ \citenamefont
  {Cortijo}(2016)}]{Alberto2016}%
  \BibitemOpen
  \bibfield  {author} {\bibinfo {author} {\bibfnamefont {Y.}~\bibnamefont
  {Ferreiros}}\ and\ \bibinfo {author} {\bibfnamefont {A.}~\bibnamefont
  {Cortijo}},\ }\href {\doibase 10.1103/PhysRevB.93.195154} {\bibfield
  {journal} {\bibinfo  {journal} {Phys. Rev. B}\ }\textbf {\bibinfo {volume}
  {93}},\ \bibinfo {pages} {195154} (\bibinfo {year} {2016})}\BibitemShut
  {NoStop}%
\bibitem [{\citenamefont {Kajita}\ \emph {et~al.}(1992)\citenamefont {Kajita},
  \citenamefont {Ojiro}, \citenamefont {Fujii}, \citenamefont {Nishio},
  \citenamefont {Kobayashi}, \citenamefont {Kobayashi},\ and\ \citenamefont
  {Kato}}]{Kajita1992}%
  \BibitemOpen
  \bibfield  {author} {\bibinfo {author} {\bibfnamefont {K.}~\bibnamefont
  {Kajita}}, \bibinfo {author} {\bibfnamefont {T.}~\bibnamefont {Ojiro}},
  \bibinfo {author} {\bibfnamefont {H.}~\bibnamefont {Fujii}}, \bibinfo
  {author} {\bibfnamefont {Y.}~\bibnamefont {Nishio}}, \bibinfo {author}
  {\bibfnamefont {H.}~\bibnamefont {Kobayashi}}, \bibinfo {author}
  {\bibfnamefont {A.}~\bibnamefont {Kobayashi}}, \ and\ \bibinfo {author}
  {\bibfnamefont {R.}~\bibnamefont {Kato}},\ }\href {\doibase
  10.1143/JPSJ.61.23} {\bibfield  {journal} {\bibinfo  {journal} {Journal of
  the Physical Society of Japan}\ }\textbf {\bibinfo {volume} {61}},\ \bibinfo
  {pages} {23} (\bibinfo {year} {1992})}\BibitemShut {NoStop}%
\bibitem [{\citenamefont {Katayama}\ \emph {et~al.}(2006)\citenamefont
  {Katayama}, \citenamefont {Kobayashi},\ and\ \citenamefont
  {Suzumura}}]{Suzumura2006}%
  \BibitemOpen
  \bibfield  {author} {\bibinfo {author} {\bibfnamefont {S.}~\bibnamefont
  {Katayama}}, \bibinfo {author} {\bibfnamefont {A.}~\bibnamefont {Kobayashi}},
  \ and\ \bibinfo {author} {\bibfnamefont {Y.}~\bibnamefont {Suzumura}},\
  }\href {\doibase 10.1143/JPSJ.75.054705} {\bibfield  {journal} {\bibinfo
  {journal} {J. Phys. Soc. Jpn.}\ }\textbf {\bibinfo {volume} {75}},\ \bibinfo
  {pages} {054705} (\bibinfo {year} {2006})}\BibitemShut {NoStop}%
\bibitem [{\citenamefont {Kajita}\ \emph {et~al.}(2014)\citenamefont {Kajita},
  \citenamefont {Nishio}, \citenamefont {Tajima}, \citenamefont {Suzumura},\
  and\ \citenamefont {Kobayashi}}]{Suzumura2014}%
  \BibitemOpen
  \bibfield  {author} {\bibinfo {author} {\bibfnamefont {K.}~\bibnamefont
  {Kajita}}, \bibinfo {author} {\bibfnamefont {Y.}~\bibnamefont {Nishio}},
  \bibinfo {author} {\bibfnamefont {N.}~\bibnamefont {Tajima}}, \bibinfo
  {author} {\bibfnamefont {Y.}~\bibnamefont {Suzumura}}, \ and\ \bibinfo
  {author} {\bibfnamefont {A.}~\bibnamefont {Kobayashi}},\ }\href {\doibase
  10.7566/JPSJ.83.072002} {\bibfield  {journal} {\bibinfo  {journal} {Journal
  of the Physical Society of Japan}\ }\textbf {\bibinfo {volume} {83}},\
  \bibinfo {pages} {072002} (\bibinfo {year} {2014})}\BibitemShut {NoStop}%
\bibitem [{\citenamefont {Zhou}\ \emph {et~al.}(2014)\citenamefont {Zhou},
  \citenamefont {Dong}, \citenamefont {Oganov}, \citenamefont {Zhu},
  \citenamefont {Tian},\ and\ \citenamefont {Wang}}]{Zhou2014}%
  \BibitemOpen
  \bibfield  {author} {\bibinfo {author} {\bibfnamefont {X.-F.}\ \bibnamefont
  {Zhou}}, \bibinfo {author} {\bibfnamefont {X.}~\bibnamefont {Dong}}, \bibinfo
  {author} {\bibfnamefont {A.~R.}\ \bibnamefont {Oganov}}, \bibinfo {author}
  {\bibfnamefont {Q.}~\bibnamefont {Zhu}}, \bibinfo {author} {\bibfnamefont
  {Y.}~\bibnamefont {Tian}}, \ and\ \bibinfo {author} {\bibfnamefont {H.-T.}\
  \bibnamefont {Wang}},\ }\href {\doibase 10.1103/PhysRevLett.112.085502}
  {\bibfield  {journal} {\bibinfo  {journal} {Phys. Rev. Lett.}\ }\textbf
  {\bibinfo {volume} {112}},\ \bibinfo {pages} {085502} (\bibinfo {year}
  {2014})}\BibitemShut {NoStop}%
\bibitem [{\citenamefont {Zabolotskiy}\ and\ \citenamefont
  {Lozovik}(2016)}]{Zabolotskiy2016}%
  \BibitemOpen
  \bibfield  {author} {\bibinfo {author} {\bibfnamefont {A.~D.}\ \bibnamefont
  {Zabolotskiy}}\ and\ \bibinfo {author} {\bibfnamefont {Y.~E.}\ \bibnamefont
  {Lozovik}},\ }\href {\doibase 10.1103/PhysRevB.94.165403} {\bibfield
  {journal} {\bibinfo  {journal} {Phys. Rev. B}\ }\textbf {\bibinfo {volume}
  {94}},\ \bibinfo {pages} {165403} (\bibinfo {year} {2016})}\BibitemShut
  {NoStop}%
\bibitem [{\citenamefont {Lopez-Bezanilla}\ and\ \citenamefont
  {Littlewood}(2016)}]{Lopez2016}%
  \BibitemOpen
  \bibfield  {author} {\bibinfo {author} {\bibfnamefont {A.}~\bibnamefont
  {Lopez-Bezanilla}}\ and\ \bibinfo {author} {\bibfnamefont {P.~B.}\
  \bibnamefont {Littlewood}},\ }\href {\doibase 10.1103/PhysRevB.93.241405}
  {\bibfield  {journal} {\bibinfo  {journal} {Phys. Rev. B}\ }\textbf {\bibinfo
  {volume} {93}},\ \bibinfo {pages} {241405} (\bibinfo {year}
  {2016})}\BibitemShut {NoStop}%
\bibitem [{\citenamefont {Feng}\ \emph {et~al.}(2017)\citenamefont {Feng},
  \citenamefont {Sugino}, \citenamefont {Liu}, \citenamefont {Zhang},
  \citenamefont {Yukawa}, \citenamefont {Kawamura}, \citenamefont {Iimori},
  \citenamefont {Kim}, \citenamefont {Hasegawa}, \citenamefont {Li},
  \citenamefont {Chen}, \citenamefont {Wu}, \citenamefont {Kumigashira},
  \citenamefont {Komori}, \citenamefont {Chiang}, \citenamefont {Meng},\ and\
  \citenamefont {Matsuda}}]{Feng2017}%
  \BibitemOpen
  \bibfield  {author} {\bibinfo {author} {\bibfnamefont {B.}~\bibnamefont
  {Feng}}, \bibinfo {author} {\bibfnamefont {O.}~\bibnamefont {Sugino}},
  \bibinfo {author} {\bibfnamefont {R.-Y.}\ \bibnamefont {Liu}}, \bibinfo
  {author} {\bibfnamefont {J.}~\bibnamefont {Zhang}}, \bibinfo {author}
  {\bibfnamefont {R.}~\bibnamefont {Yukawa}}, \bibinfo {author} {\bibfnamefont
  {M.}~\bibnamefont {Kawamura}}, \bibinfo {author} {\bibfnamefont
  {T.}~\bibnamefont {Iimori}}, \bibinfo {author} {\bibfnamefont
  {H.}~\bibnamefont {Kim}}, \bibinfo {author} {\bibfnamefont {Y.}~\bibnamefont
  {Hasegawa}}, \bibinfo {author} {\bibfnamefont {H.}~\bibnamefont {Li}},
  \bibinfo {author} {\bibfnamefont {L.}~\bibnamefont {Chen}}, \bibinfo {author}
  {\bibfnamefont {K.}~\bibnamefont {Wu}}, \bibinfo {author} {\bibfnamefont
  {H.}~\bibnamefont {Kumigashira}}, \bibinfo {author} {\bibfnamefont
  {F.}~\bibnamefont {Komori}}, \bibinfo {author} {\bibfnamefont {T.-C.}\
  \bibnamefont {Chiang}}, \bibinfo {author} {\bibfnamefont {S.}~\bibnamefont
  {Meng}}, \ and\ \bibinfo {author} {\bibfnamefont {I.}~\bibnamefont
  {Matsuda}},\ }\href {\doibase 10.1103/PhysRevLett.118.096401} {\bibfield
  {journal} {\bibinfo  {journal} {Phys. Rev. Lett.}\ }\textbf {\bibinfo
  {volume} {118}},\ \bibinfo {pages} {096401} (\bibinfo {year}
  {2017})}\BibitemShut {NoStop}%
\bibitem [{\citenamefont {Tao}\ and\ \citenamefont
  {Tsymbal}(2018)}]{Tsymbal2018}%
  \BibitemOpen
  \bibfield  {author} {\bibinfo {author} {\bibfnamefont {L.~L.}\ \bibnamefont
  {Tao}}\ and\ \bibinfo {author} {\bibfnamefont {E.~Y.}\ \bibnamefont
  {Tsymbal}},\ }\href {\doibase 10.1103/PhysRevB.98.121102} {\bibfield
  {journal} {\bibinfo  {journal} {Phys. Rev. B}\ }\textbf {\bibinfo {volume}
  {98}},\ \bibinfo {pages} {121102} (\bibinfo {year} {2018})}\BibitemShut
  {NoStop}%
\bibitem [{\citenamefont {Goerbig}\ \emph {et~al.}(2008)\citenamefont
  {Goerbig}, \citenamefont {Fuchs}, \citenamefont {Montambaux},\ and\
  \citenamefont {Pi\'echon}}]{Goerbig2008}%
  \BibitemOpen
  \bibfield  {author} {\bibinfo {author} {\bibfnamefont {M.~O.}\ \bibnamefont
  {Goerbig}}, \bibinfo {author} {\bibfnamefont {J.-N.}\ \bibnamefont {Fuchs}},
  \bibinfo {author} {\bibfnamefont {G.}~\bibnamefont {Montambaux}}, \ and\
  \bibinfo {author} {\bibfnamefont {F.}~\bibnamefont {Pi\'echon}},\ }\href
  {\doibase 10.1103/PhysRevB.78.045415} {\bibfield  {journal} {\bibinfo
  {journal} {Phys. Rev. B}\ }\textbf {\bibinfo {volume} {78}},\ \bibinfo
  {pages} {045415} (\bibinfo {year} {2008})}\BibitemShut {NoStop}%
\bibitem [{\citenamefont {Lu}\ \emph {et~al.}(2016)\citenamefont {Lu},
  \citenamefont {Cuamba}, \citenamefont {Lin}, \citenamefont {Hao},
  \citenamefont {Wang}, \citenamefont {Li}, \citenamefont {Zhao},\ and\
  \citenamefont {Ting}}]{Lu2016}%
  \BibitemOpen
  \bibfield  {author} {\bibinfo {author} {\bibfnamefont {H.-Y.}\ \bibnamefont
  {Lu}}, \bibinfo {author} {\bibfnamefont {A.~S.}\ \bibnamefont {Cuamba}},
  \bibinfo {author} {\bibfnamefont {S.-Y.}\ \bibnamefont {Lin}}, \bibinfo
  {author} {\bibfnamefont {L.}~\bibnamefont {Hao}}, \bibinfo {author}
  {\bibfnamefont {R.}~\bibnamefont {Wang}}, \bibinfo {author} {\bibfnamefont
  {H.}~\bibnamefont {Li}}, \bibinfo {author} {\bibfnamefont {Y.}~\bibnamefont
  {Zhao}}, \ and\ \bibinfo {author} {\bibfnamefont {C.~S.}\ \bibnamefont
  {Ting}},\ }\href {\doibase 10.1103/PhysRevB.94.195423} {\bibfield  {journal}
  {\bibinfo  {journal} {Phys. Rev. B}\ }\textbf {\bibinfo {volume} {94}},\
  \bibinfo {pages} {195423} (\bibinfo {year} {2016})}\BibitemShut {NoStop}%
\bibitem [{\citenamefont {Yan}\ \emph {et~al.}(2017)\citenamefont {Yan},
  \citenamefont {Huang}, \citenamefont {Zhang}, \citenamefont {Wang},
  \citenamefont {Yao}, \citenamefont {Deng}, \citenamefont {Wan}, \citenamefont
  {Zhang}, \citenamefont {Arita}, \citenamefont {Yang}, \citenamefont {Sun},
  \citenamefont {Yao}, \citenamefont {Wu}, \citenamefont {Fan}, \citenamefont
  {Duan},\ and\ \citenamefont {Zhou}}]{Yan2017}%
  \BibitemOpen
  \bibfield  {author} {\bibinfo {author} {\bibfnamefont {M.}~\bibnamefont
  {Yan}}, \bibinfo {author} {\bibfnamefont {H.}~\bibnamefont {Huang}}, \bibinfo
  {author} {\bibfnamefont {K.}~\bibnamefont {Zhang}}, \bibinfo {author}
  {\bibfnamefont {E.}~\bibnamefont {Wang}}, \bibinfo {author} {\bibfnamefont
  {W.}~\bibnamefont {Yao}}, \bibinfo {author} {\bibfnamefont {K.}~\bibnamefont
  {Deng}}, \bibinfo {author} {\bibfnamefont {G.}~\bibnamefont {Wan}}, \bibinfo
  {author} {\bibfnamefont {H.}~\bibnamefont {Zhang}}, \bibinfo {author}
  {\bibfnamefont {M.}~\bibnamefont {Arita}}, \bibinfo {author} {\bibfnamefont
  {H.}~\bibnamefont {Yang}}, \bibinfo {author} {\bibfnamefont {Z.}~\bibnamefont
  {Sun}}, \bibinfo {author} {\bibfnamefont {H.}~\bibnamefont {Yao}}, \bibinfo
  {author} {\bibfnamefont {Y.}~\bibnamefont {Wu}}, \bibinfo {author}
  {\bibfnamefont {S.}~\bibnamefont {Fan}}, \bibinfo {author} {\bibfnamefont
  {W.}~\bibnamefont {Duan}}, \ and\ \bibinfo {author} {\bibfnamefont
  {S.}~\bibnamefont {Zhou}},\ }\href {\doibase 10.1038/s41467-017-00280-6}
  {\bibfield  {journal} {\bibinfo  {journal} {Nature Communications}\ }\textbf
  {\bibinfo {volume} {8}},\ \bibinfo {pages} {257} (\bibinfo {year}
  {2017})}\BibitemShut {NoStop}%
\bibitem [{\citenamefont {Noh}\ \emph {et~al.}(2017)\citenamefont {Noh},
  \citenamefont {Jeong}, \citenamefont {Cho}, \citenamefont {Kim},
  \citenamefont {Min},\ and\ \citenamefont {Park}}]{Noh}%
  \BibitemOpen
  \bibfield  {author} {\bibinfo {author} {\bibfnamefont {H.-J.}\ \bibnamefont
  {Noh}}, \bibinfo {author} {\bibfnamefont {J.}~\bibnamefont {Jeong}}, \bibinfo
  {author} {\bibfnamefont {E.-J.}\ \bibnamefont {Cho}}, \bibinfo {author}
  {\bibfnamefont {K.}~\bibnamefont {Kim}}, \bibinfo {author} {\bibfnamefont
  {B.~I.}\ \bibnamefont {Min}}, \ and\ \bibinfo {author} {\bibfnamefont
  {B.-G.}\ \bibnamefont {Park}},\ }\href
  {https://link.aps.org/doi/10.1103/PhysRevLett.119.016401} {\bibfield
  {journal} {\bibinfo  {journal} {Phys. Rev. Lett.}\ }\textbf {\bibinfo
  {volume} {119}},\ \bibinfo {pages} {016401} (\bibinfo {year}
  {2017})}\BibitemShut {NoStop}%
\bibitem [{\citenamefont {Soluyanov}\ \emph {et~al.}(2015)\citenamefont
  {Soluyanov}, \citenamefont {Gresch}, \citenamefont {Wang}, \citenamefont
  {Wu}, \citenamefont {Troyer}, \citenamefont {Dai},\ and\ \citenamefont
  {Bernevig}}]{Soluyanov2015}%
  \BibitemOpen
  \bibfield  {author} {\bibinfo {author} {\bibfnamefont {A.~A.}\ \bibnamefont
  {Soluyanov}}, \bibinfo {author} {\bibfnamefont {D.}~\bibnamefont {Gresch}},
  \bibinfo {author} {\bibfnamefont {Z.}~\bibnamefont {Wang}}, \bibinfo {author}
  {\bibfnamefont {Q.}~\bibnamefont {Wu}}, \bibinfo {author} {\bibfnamefont
  {M.}~\bibnamefont {Troyer}}, \bibinfo {author} {\bibfnamefont
  {X.}~\bibnamefont {Dai}}, \ and\ \bibinfo {author} {\bibfnamefont {B.~A.}\
  \bibnamefont {Bernevig}},\ }\href {\doibase 10.1038/nature15768} {\bibfield
  {journal} {\bibinfo  {journal} {Nature}\ }\textbf {\bibinfo {volume} {527}},\
  \bibinfo {pages} {495} (\bibinfo {year} {2015})}\BibitemShut {NoStop}%
\bibitem [{\citenamefont {Wang}\ \emph {et~al.}(2016)\citenamefont {Wang},
  \citenamefont {Gresch}, \citenamefont {Soluyanov}, \citenamefont {Xie},
  \citenamefont {Kushwaha}, \citenamefont {Dai}, \citenamefont {Troyer},
  \citenamefont {Cava},\ and\ \citenamefont {Bernevig}}]{Zhijun-2016}%
  \BibitemOpen
  \bibfield  {author} {\bibinfo {author} {\bibfnamefont {Z.}~\bibnamefont
  {Wang}}, \bibinfo {author} {\bibfnamefont {D.}~\bibnamefont {Gresch}},
  \bibinfo {author} {\bibfnamefont {A.~A.}\ \bibnamefont {Soluyanov}}, \bibinfo
  {author} {\bibfnamefont {W.}~\bibnamefont {Xie}}, \bibinfo {author}
  {\bibfnamefont {S.}~\bibnamefont {Kushwaha}}, \bibinfo {author}
  {\bibfnamefont {X.}~\bibnamefont {Dai}}, \bibinfo {author} {\bibfnamefont
  {M.}~\bibnamefont {Troyer}}, \bibinfo {author} {\bibfnamefont {R.~J.}\
  \bibnamefont {Cava}}, \ and\ \bibinfo {author} {\bibfnamefont {B.~A.}\
  \bibnamefont {Bernevig}},\ }\href {\doibase 10.1103/PhysRevLett.117.056805}
  {\bibfield  {journal} {\bibinfo  {journal} {Phys. Rev. Lett.}\ }\textbf
  {\bibinfo {volume} {117}},\ \bibinfo {pages} {056805} (\bibinfo {year}
  {2016})}\BibitemShut {NoStop}%
\bibitem [{\citenamefont {Rostamzadeh}\ \emph {et~al.}(2019)\citenamefont
  {Rostamzadeh}, \citenamefont {Adagideli},\ and\ \citenamefont
  {Goerbig}}]{Adagideli2019}%
  \BibitemOpen
  \bibfield  {author} {\bibinfo {author} {\bibfnamefont {S.}~\bibnamefont
  {Rostamzadeh}}, \bibinfo {author} {\bibfnamefont {i.~d. I. m.~c.}\
  \bibnamefont {Adagideli}}, \ and\ \bibinfo {author} {\bibfnamefont {M.~O.}\
  \bibnamefont {Goerbig}},\ }\href {\doibase 10.1103/PhysRevB.100.075438}
  {\bibfield  {journal} {\bibinfo  {journal} {Phys. Rev. B}\ }\textbf {\bibinfo
  {volume} {100}},\ \bibinfo {pages} {075438} (\bibinfo {year}
  {2019})}\BibitemShut {NoStop}%
\bibitem [{\citenamefont {Zhang}\ \emph {et~al.}(2018)\citenamefont {Zhang},
  \citenamefont {Zhang}, \citenamefont {Xia}, \citenamefont {Gao},\ and\
  \citenamefont {Zhang}}]{Zhang2018}%
  \BibitemOpen
  \bibfield  {author} {\bibinfo {author} {\bibfnamefont {K.}~\bibnamefont
  {Zhang}}, \bibinfo {author} {\bibfnamefont {E.}~\bibnamefont {Zhang}},
  \bibinfo {author} {\bibfnamefont {M.}~\bibnamefont {Xia}}, \bibinfo {author}
  {\bibfnamefont {P.}~\bibnamefont {Gao}}, \ and\ \bibinfo {author}
  {\bibfnamefont {S.}~\bibnamefont {Zhang}},\ }\href
  {http://www.sciencedirect.com/science/article/pii/S0003491618301118}
  {\bibfield  {journal} {\bibinfo  {journal} {Annals of Physics}\ }\textbf
  {\bibinfo {volume} {394}},\ \bibinfo {pages} {1 } (\bibinfo {year}
  {2018})}\BibitemShut {NoStop}%
\bibitem [{\citenamefont {Jalali-Mola}\ and\ \citenamefont
  {Jafari}(2019{\natexlab{b}})}]{SaharSSP}%
  \BibitemOpen
  \bibfield  {author} {\bibinfo {author} {\bibfnamefont {Z.}~\bibnamefont
  {Jalali-Mola}}\ and\ \bibinfo {author} {\bibfnamefont {S.~A.}\ \bibnamefont
  {Jafari}},\ }\href {\doibase 10.1103/PhysRevB.100.205413} {\bibfield
  {journal} {\bibinfo  {journal} {Phys. Rev. B}\ }\textbf {\bibinfo {volume}
  {100}},\ \bibinfo {pages} {205413} (\bibinfo {year}
  {2019}{\natexlab{b}})}\BibitemShut {NoStop}%
\bibitem [{\citenamefont {Jalali-Mola}\ and\ \citenamefont
  {Jafari}(2018{\natexlab{a}})}]{SaharTilt1}%
  \BibitemOpen
  \bibfield  {author} {\bibinfo {author} {\bibfnamefont {Z.}~\bibnamefont
  {Jalali-Mola}}\ and\ \bibinfo {author} {\bibfnamefont {S.~A.}\ \bibnamefont
  {Jafari}},\ }\href {\doibase 10.1103/PhysRevB.98.195415} {\bibfield
  {journal} {\bibinfo  {journal} {Phys. Rev. B}\ }\textbf {\bibinfo {volume}
  {98}},\ \bibinfo {pages} {195415} (\bibinfo {year}
  {2018}{\natexlab{a}})}\BibitemShut {NoStop}%
\bibitem [{\citenamefont {Jalali-Mola}\ and\ \citenamefont
  {Jafari}(2018{\natexlab{b}})}]{SaharTilt2}%
  \BibitemOpen
  \bibfield  {author} {\bibinfo {author} {\bibfnamefont {Z.}~\bibnamefont
  {Jalali-Mola}}\ and\ \bibinfo {author} {\bibfnamefont {S.~A.}\ \bibnamefont
  {Jafari}},\ }\href {\doibase 10.1103/PhysRevB.98.235430} {\bibfield
  {journal} {\bibinfo  {journal} {Phys. Rev. B}\ }\textbf {\bibinfo {volume}
  {98}},\ \bibinfo {pages} {235430} (\bibinfo {year}
  {2018}{\natexlab{b}})}\BibitemShut {NoStop}%
\bibitem [{\citenamefont {S\'ari}\ \emph {et~al.}(2014)\citenamefont {S\'ari},
  \citenamefont {T\ifmmode~\mbox{\H{o}}\else \H{o}\fi{}ke},\ and\ \citenamefont
  {Goerbig}}]{Goerbig2014}%
  \BibitemOpen
  \bibfield  {author} {\bibinfo {author} {\bibfnamefont {J.}~\bibnamefont
  {S\'ari}}, \bibinfo {author} {\bibfnamefont {C.}~\bibnamefont
  {T\ifmmode~\mbox{\H{o}}\else \H{o}\fi{}ke}}, \ and\ \bibinfo {author}
  {\bibfnamefont {M.~O.}\ \bibnamefont {Goerbig}},\ }\href {\doibase
  10.1103/PhysRevB.90.155446} {\bibfield  {journal} {\bibinfo  {journal} {Phys.
  Rev. B}\ }\textbf {\bibinfo {volume} {90}},\ \bibinfo {pages} {155446}
  (\bibinfo {year} {2014})}\BibitemShut {NoStop}%
\bibitem [{\citenamefont {Zyuzin}\ and\ \citenamefont
  {Tiwari}(2016)}]{Zyuzin2016}%
  \BibitemOpen
  \bibfield  {author} {\bibinfo {author} {\bibfnamefont {A.~A.}\ \bibnamefont
  {Zyuzin}}\ and\ \bibinfo {author} {\bibfnamefont {R.~P.}\ \bibnamefont
  {Tiwari}},\ }\href {\doibase 10.1134/S002136401611014X} {\bibfield  {journal}
  {\bibinfo  {journal} {JETP Letters}\ }\textbf {\bibinfo {volume} {103}},\
  \bibinfo {pages} {717} (\bibinfo {year} {2016})}\BibitemShut {NoStop}%
\bibitem [{\citenamefont {Zhao}\ and\ \citenamefont {Wang}(2019)}]{Zhao2019}%
  \BibitemOpen
  \bibfield  {author} {\bibinfo {author} {\bibfnamefont {P.-L.}\ \bibnamefont
  {Zhao}}\ and\ \bibinfo {author} {\bibfnamefont {A.-M.}\ \bibnamefont
  {Wang}},\ }\href {\doibase 10.1103/PhysRevB.100.125138} {\bibfield  {journal}
  {\bibinfo  {journal} {Phys. Rev. B}\ }\textbf {\bibinfo {volume} {100}},\
  \bibinfo {pages} {125138} (\bibinfo {year} {2019})}\BibitemShut {NoStop}%
\bibitem [{\citenamefont {Lee}\ and\ \citenamefont {Lee}(2018)}]{Lee2018}%
  \BibitemOpen
  \bibfield  {author} {\bibinfo {author} {\bibfnamefont {Y.-W.}\ \bibnamefont
  {Lee}}\ and\ \bibinfo {author} {\bibfnamefont {Y.-L.}\ \bibnamefont {Lee}},\
  }\href {\doibase 10.1103/PhysRevB.97.035141} {\bibfield  {journal} {\bibinfo
  {journal} {Phys. Rev. B}\ }\textbf {\bibinfo {volume} {97}},\ \bibinfo
  {pages} {035141} (\bibinfo {year} {2018})}\BibitemShut {NoStop}%
\bibitem [{\citenamefont {Faraei}\ and\ \citenamefont
  {Jafari}(2019{\natexlab{a}})}]{FaraeiPAR}%
  \BibitemOpen
  \bibfield  {author} {\bibinfo {author} {\bibfnamefont {Z.}~\bibnamefont
  {Faraei}}\ and\ \bibinfo {author} {\bibfnamefont {S.~A.}\ \bibnamefont
  {Jafari}},\ }\href {\doibase 10.1103/PhysRevB.100.245436} {\bibfield
  {journal} {\bibinfo  {journal} {Phys. Rev. B}\ }\textbf {\bibinfo {volume}
  {100}},\ \bibinfo {pages} {245436} (\bibinfo {year}
  {2019}{\natexlab{a}})}\BibitemShut {NoStop}%
\bibitem [{\citenamefont {Faraei}\ and\ \citenamefont
  {Jafari}(2019{\natexlab{b}})}]{FaraeiCharged}%
  \BibitemOpen
  \bibfield  {author} {\bibinfo {author} {\bibfnamefont {Z.}~\bibnamefont
  {Faraei}}\ and\ \bibinfo {author} {\bibfnamefont {S.}~\bibnamefont
  {Jafari}},\ }\href {https://arxiv.org/abs/1912.06355} {\bibfield  {journal}
  {\bibinfo  {journal} {arXiv:1912.06355}\ } (\bibinfo {year}
  {2019}{\natexlab{b}})}\BibitemShut {NoStop}%
\bibitem [{\citenamefont {Jafari}(2019)}]{JafariEB}%
  \BibitemOpen
  \bibfield  {author} {\bibinfo {author} {\bibfnamefont {S.~A.}\ \bibnamefont
  {Jafari}},\ }\href {\doibase 10.1103/PhysRevB.100.045144} {\bibfield
  {journal} {\bibinfo  {journal} {Phys. Rev. B}\ }\textbf {\bibinfo {volume}
  {100}},\ \bibinfo {pages} {045144} (\bibinfo {year} {2019})}\BibitemShut
  {NoStop}%
\bibitem [{\citenamefont {Jalali-Mola}\ and\ \citenamefont
  {Jafari}(2019{\natexlab{c}})}]{SaharCovariance}%
  \BibitemOpen
  \bibfield  {author} {\bibinfo {author} {\bibfnamefont {Z.}~\bibnamefont
  {Jalali-Mola}}\ and\ \bibinfo {author} {\bibfnamefont {S.~A.}\ \bibnamefont
  {Jafari}},\ }\href {\doibase 10.1103/PhysRevB.100.075113} {\bibfield
  {journal} {\bibinfo  {journal} {Phys. Rev. B}\ }\textbf {\bibinfo {volume}
  {100}},\ \bibinfo {pages} {075113} (\bibinfo {year}
  {2019}{\natexlab{c}})}\BibitemShut {NoStop}%
\bibitem [{\citenamefont {Volovik}(2016)}]{Volovik2016black}%
  \BibitemOpen
  \bibfield  {author} {\bibinfo {author} {\bibfnamefont {G.~E.}\ \bibnamefont
  {Volovik}},\ }\href {https://doi.org/10.1134/S0021364016210050} {\bibfield
  {journal} {\bibinfo  {journal} {JETP letters}\ }\textbf {\bibinfo {volume}
  {104}},\ \bibinfo {pages} {645} (\bibinfo {year} {2016})}\BibitemShut
  {NoStop}%
\bibitem [{\citenamefont {Zhang}\ and\ \citenamefont
  {Volovik}(2017)}]{Volovik2017lifshitz}%
  \BibitemOpen
  \bibfield  {author} {\bibinfo {author} {\bibfnamefont {K.}~\bibnamefont
  {Zhang}}\ and\ \bibinfo {author} {\bibfnamefont {G.~E.}\ \bibnamefont
  {Volovik}},\ }\href {\doibase 10.1134/S0021364017080094} {\bibfield
  {journal} {\bibinfo  {journal} {JETP Letters}\ }\textbf {\bibinfo {volume}
  {105}},\ \bibinfo {pages} {519} (\bibinfo {year} {2017})}\BibitemShut
  {NoStop}%
\bibitem [{\citenamefont {Volovik}(2018)}]{Volovik2018}%
  \BibitemOpen
  \bibfield  {author} {\bibinfo {author} {\bibfnamefont {G.~E.}\ \bibnamefont
  {Volovik}},\ }\href {\doibase 10.3367/ufne.2017.01.038218} {\bibfield
  {journal} {\bibinfo  {journal} {Physics-Uspekhi}\ }\textbf {\bibinfo {volume}
  {61}},\ \bibinfo {pages} {89} (\bibinfo {year} {2018})}\BibitemShut {NoStop}%
\bibitem [{\citenamefont {Farajollahpour}\ \emph {et~al.}(2019)\citenamefont
  {Farajollahpour}, \citenamefont {Faraei},\ and\ \citenamefont
  {Jafari}}]{TohidBorophene}%
  \BibitemOpen
  \bibfield  {author} {\bibinfo {author} {\bibfnamefont {T.}~\bibnamefont
  {Farajollahpour}}, \bibinfo {author} {\bibfnamefont {Z.}~\bibnamefont
  {Faraei}}, \ and\ \bibinfo {author} {\bibfnamefont {S.~A.}\ \bibnamefont
  {Jafari}},\ }\href {\doibase 10.1103/PhysRevB.99.235150} {\bibfield
  {journal} {\bibinfo  {journal} {Phys. Rev. B}\ }\textbf {\bibinfo {volume}
  {99}},\ \bibinfo {pages} {235150} (\bibinfo {year} {2019})}\BibitemShut
  {NoStop}%
\bibitem [{\citenamefont {Hashimoto}\ and\ \citenamefont
  {Matsuo}(2019)}]{Hashimoto2019}%
  \BibitemOpen
  \bibfield  {author} {\bibinfo {author} {\bibfnamefont {K.}~\bibnamefont
  {Hashimoto}}\ and\ \bibinfo {author} {\bibfnamefont {Y.}~\bibnamefont
  {Matsuo}},\ }\href@noop {} {\enquote {\bibinfo {title} {Escape from black
  holes in materials: Type ii weyl semimetals and generic edge states},}\ }
  (\bibinfo {year} {2019}),\ \Eprint {http://arxiv.org/abs/1911.04675}
  {arXiv:1911.04675 [cond-mat.mes-hall]} \BibitemShut {NoStop}%
\bibitem [{\citenamefont {Kedem}\ \emph {et~al.}(2020)\citenamefont {Kedem},
  \citenamefont {Bergholtz},\ and\ \citenamefont {Wilczek}}]{Wilczek2020}%
  \BibitemOpen
  \bibfield  {author} {\bibinfo {author} {\bibfnamefont {Y.}~\bibnamefont
  {Kedem}}, \bibinfo {author} {\bibfnamefont {E.~J.}\ \bibnamefont
  {Bergholtz}}, \ and\ \bibinfo {author} {\bibfnamefont {F.}~\bibnamefont
  {Wilczek}},\ }\href@noop {} {\  (\bibinfo {year} {2020})},\ \Eprint
  {http://arxiv.org/abs/2001.02625} {arXiv:2001.02625 [cond-mat.mes-hall]}
  \BibitemShut {NoStop}%
\bibitem [{\citenamefont {Nakayama}(1974)}]{Nakayama1974}%
  \BibitemOpen
  \bibfield  {author} {\bibinfo {author} {\bibfnamefont {M.}~\bibnamefont
  {Nakayama}},\ }\href {\doibase 10.1143/JPSJ.36.393} {\bibfield  {journal}
  {\bibinfo  {journal} {Journal of the Physical Society of Japan}\ }\textbf
  {\bibinfo {volume} {36}},\ \bibinfo {pages} {393} (\bibinfo {year}
  {1974})}\BibitemShut {NoStop}%
\bibitem [{\citenamefont {Maier}(2007)}]{Maier-Book}%
  \BibitemOpen
  \bibfield  {author} {\bibinfo {author} {\bibfnamefont {S.~A.}\ \bibnamefont
  {Maier}},\ }\href@noop {} {\emph {\bibinfo {title} {Plasmonics: fundamentals
  and applications}}}\ (\bibinfo  {publisher} {Springer Science \& Business
  Media, New York},\ \bibinfo {year} {2007})\BibitemShut {NoStop}%
\bibitem [{\citenamefont {Fetter}(1973)}]{Fetter1973}%
  \BibitemOpen
  \bibfield  {author} {\bibinfo {author} {\bibfnamefont {A.~L.}\ \bibnamefont
  {Fetter}},\ }\href
  {https://www.sciencedirect.com/science/article/pii/0003491673901619}
  {\bibfield  {journal} {\bibinfo  {journal} {Ann. Phys.}\ }\textbf {\bibinfo
  {volume} {81}},\ \bibinfo {pages} {367} (\bibinfo {year} {1973})}\BibitemShut
  {NoStop}%
\bibitem [{\citenamefont {Lucas}\ and\ \citenamefont
  {Das~Sarma}(2018)}]{Lucas2018}%
  \BibitemOpen
  \bibfield  {author} {\bibinfo {author} {\bibfnamefont {A.}~\bibnamefont
  {Lucas}}\ and\ \bibinfo {author} {\bibfnamefont {S.}~\bibnamefont
  {Das~Sarma}},\ }\href {\doibase 10.1103/PhysRevB.97.115449} {\bibfield
  {journal} {\bibinfo  {journal} {Phys. Rev. B}\ }\textbf {\bibinfo {volume}
  {97}},\ \bibinfo {pages} {115449} (\bibinfo {year} {2018})}\BibitemShut
  {NoStop}%
\bibitem [{\citenamefont {Okamoto}(2006)}]{Okamoto-Book}%
  \BibitemOpen
  \bibfield  {author} {\bibinfo {author} {\bibfnamefont {K.}~\bibnamefont
  {Okamoto}},\ }\href@noop {} {\emph {\bibinfo {title} {Fundamentals of optical
  waveguides}}},\ \bibinfo {edition} {second edition}\ ed.\ (\bibinfo
  {publisher} {Academic press,United States},\ \bibinfo {year}
  {2006})\BibitemShut {NoStop}%
\bibitem [{\citenamefont {Huang}\ \emph {et~al.}(2018)\citenamefont {Huang},
  \citenamefont {Jin},\ and\ \citenamefont {Liu}}]{Feng2018}%
  \BibitemOpen
  \bibfield  {author} {\bibinfo {author} {\bibfnamefont {H.}~\bibnamefont
  {Huang}}, \bibinfo {author} {\bibfnamefont {K.-H.}\ \bibnamefont {Jin}}, \
  and\ \bibinfo {author} {\bibfnamefont {F.}~\bibnamefont {Liu}},\ }\href
  {\doibase 10.1103/PhysRevB.98.121110} {\bibfield  {journal} {\bibinfo
  {journal} {Phys. Rev. B}\ }\textbf {\bibinfo {volume} {98}},\ \bibinfo
  {pages} {121110} (\bibinfo {year} {2018})}\BibitemShut {NoStop}%
\bibitem [{\citenamefont {Kang}\ \emph {et~al.}(2019)\citenamefont {Kang},
  \citenamefont {Huang}, \citenamefont {Zhang}, \citenamefont {Xu},\ and\
  \citenamefont {Liu}}]{kang2019optical}%
  \BibitemOpen
  \bibfield  {author} {\bibinfo {author} {\bibfnamefont {M.}~\bibnamefont
  {Kang}}, \bibinfo {author} {\bibfnamefont {H.}~\bibnamefont {Huang}},
  \bibinfo {author} {\bibfnamefont {S.}~\bibnamefont {Zhang}}, \bibinfo
  {author} {\bibfnamefont {H.}~\bibnamefont {Xu}}, \ and\ \bibinfo {author}
  {\bibfnamefont {F.}~\bibnamefont {Liu}},\ }\href
  {https://arxiv.org/abs/1908.05049} {\bibfield  {journal} {\bibinfo  {journal}
  {arXiv preprint arXiv:1908.05049}\ } (\bibinfo {year} {2019})}\BibitemShut
  {NoStop}%
\bibitem [{\citenamefont {Liu}\ \emph {et~al.}(2018)\citenamefont {Liu},
  \citenamefont {Sun}, \citenamefont {Huang}, \citenamefont {Liu},\ and\
  \citenamefont {Meng}}]{HawkingTemperatureSun}%
  \BibitemOpen
  \bibfield  {author} {\bibinfo {author} {\bibfnamefont {H.}~\bibnamefont
  {Liu}}, \bibinfo {author} {\bibfnamefont {J.-T.}\ \bibnamefont {Sun}},
  \bibinfo {author} {\bibfnamefont {H.}~\bibnamefont {Huang}}, \bibinfo
  {author} {\bibfnamefont {F.}~\bibnamefont {Liu}}, \ and\ \bibinfo {author}
  {\bibfnamefont {S.}~\bibnamefont {Meng}},\ }\href
  {https://arxiv.org/abs/1809.00479} {\bibfield  {journal} {\bibinfo  {journal}
  {arXiv preprint arXiv:1809.00479}\ } (\bibinfo {year} {2018})}\BibitemShut
  {NoStop}%
\bibitem [{\citenamefont {Wen}(2004)}]{wen}%
  \BibitemOpen
  \bibfield  {author} {\bibinfo {author} {\bibfnamefont {X.~G.}\ \bibnamefont
  {Wen}},\ }\href@noop {} {\emph {\bibinfo {title} {Quantum Field Theory of
  Many-Body Systems}}}\ (\bibinfo  {publisher} {Oxford U. P.},\ \bibinfo {year}
  {2004})\BibitemShut {NoStop}%
\bibitem [{\citenamefont {S{\'o}lyom}(2010)}]{Solyom2010-vol3}%
  \BibitemOpen
  \bibfield  {author} {\bibinfo {author} {\bibfnamefont {J.}~\bibnamefont
  {S{\'o}lyom}},\ }\href@noop {} {\emph {\bibinfo {title} {Fundamentals of the
  Physics of Solids: Volume 3-Normal, Broken-Symmetry, and Correlated
  Systems}}},\ Vol.~\bibinfo {volume} {3}\ (\bibinfo  {publisher} {Springer
  Science \& Business Media},\ \bibinfo {year} {2010})\BibitemShut {NoStop}%
\bibitem [{\citenamefont {B.~Wunsch}\ \emph {et~al.}(2006)\citenamefont
  {B.~Wunsch}, \citenamefont {Stauber}, \citenamefont {Sols},\ and\
  \citenamefont {Guinea}}]{Wunsch2006}%
  \BibitemOpen
  \bibfield  {author} {\bibinfo {author} {\bibfnamefont {T.}~\bibnamefont
  {B.~Wunsch}}, \bibinfo {author} {\bibnamefont {Stauber}}, \bibinfo {author}
  {\bibfnamefont {F.}~\bibnamefont {Sols}}, \ and\ \bibinfo {author}
  {\bibfnamefont {F.}~\bibnamefont {Guinea}},\ }\href {\doibase
  10.1088/1367-2630/8/12/318} {\bibfield  {journal} {\bibinfo  {journal} {New
  Journal of Physics}\ }\textbf {\bibinfo {volume} {8}},\ \bibinfo {pages}
  {318} (\bibinfo {year} {2006})}\BibitemShut {NoStop}%
\bibitem [{\citenamefont {Hwang}\ and\ \citenamefont
  {Das~Sarma}(2007)}]{Hwang2007}%
  \BibitemOpen
  \bibfield  {author} {\bibinfo {author} {\bibfnamefont {E.~H.}\ \bibnamefont
  {Hwang}}\ and\ \bibinfo {author} {\bibfnamefont {S.}~\bibnamefont
  {Das~Sarma}},\ }\href {\doibase 10.1103/PhysRevB.75.205418} {\bibfield
  {journal} {\bibinfo  {journal} {Phys. Rev. B}\ }\textbf {\bibinfo {volume}
  {75}},\ \bibinfo {pages} {205418} (\bibinfo {year} {2007})}\BibitemShut
  {NoStop}%
\bibitem [{\citenamefont {Grigorenko}\ \emph {et~al.}(2012)\citenamefont
  {Grigorenko}, \citenamefont {Polini},\ and\ \citenamefont
  {Novoselov}}]{Novoselov2012}%
  \BibitemOpen
  \bibfield  {author} {\bibinfo {author} {\bibfnamefont {A.}~\bibnamefont
  {Grigorenko}}, \bibinfo {author} {\bibfnamefont {M.}~\bibnamefont {Polini}},
  \ and\ \bibinfo {author} {\bibfnamefont {K.}~\bibnamefont {Novoselov}},\
  }\href {\doibase https://doi.org/10.1038/nphoton.2012.262} {\bibfield
  {journal} {\bibinfo  {journal} {Nature photonics}\ }\textbf {\bibinfo
  {volume} {6}},\ \bibinfo {pages} {749} (\bibinfo {year} {2012})}\BibitemShut
  {NoStop}%
\bibitem [{\citenamefont {Gangadharaiah}\ \emph {et~al.}(2008)\citenamefont
  {Gangadharaiah}, \citenamefont {Farid},\ and\ \citenamefont
  {Mishchenko}}]{Mishchenko}%
  \BibitemOpen
  \bibfield  {author} {\bibinfo {author} {\bibfnamefont {S.}~\bibnamefont
  {Gangadharaiah}}, \bibinfo {author} {\bibfnamefont {A.~M.}\ \bibnamefont
  {Farid}}, \ and\ \bibinfo {author} {\bibfnamefont {E.~G.}\ \bibnamefont
  {Mishchenko}},\ }\href {\doibase 10.1103/PhysRevLett.100.166802} {\bibfield
  {journal} {\bibinfo  {journal} {Phys. Rev. Lett.}\ }\textbf {\bibinfo
  {volume} {100}},\ \bibinfo {pages} {166802} (\bibinfo {year}
  {2008})}\BibitemShut {NoStop}%
\bibitem [{\citenamefont {Vafek}(2006)}]{Vafek2006}%
  \BibitemOpen
  \bibfield  {author} {\bibinfo {author} {\bibfnamefont {O.}~\bibnamefont
  {Vafek}},\ }\href {\doibase 10.1103/PhysRevLett.97.266406} {\bibfield
  {journal} {\bibinfo  {journal} {Phys. Rev. Lett.}\ }\textbf {\bibinfo
  {volume} {97}},\ \bibinfo {pages} {266406} (\bibinfo {year}
  {2006})}\BibitemShut {NoStop}%
\bibitem [{\citenamefont {Das~Sarma}\ and\ \citenamefont
  {Li}(2013)}]{Sarma2013}%
  \BibitemOpen
  \bibfield  {author} {\bibinfo {author} {\bibfnamefont {S.}~\bibnamefont
  {Das~Sarma}}\ and\ \bibinfo {author} {\bibfnamefont {Q.}~\bibnamefont {Li}},\
  }\href {\doibase 10.1103/PhysRevB.87.235418} {\bibfield  {journal} {\bibinfo
  {journal} {Phys. Rev. B}\ }\textbf {\bibinfo {volume} {87}},\ \bibinfo
  {pages} {235418} (\bibinfo {year} {2013})}\BibitemShut {NoStop}%
\bibitem [{\citenamefont {Novoselov}\ \emph {et~al.}(2004)\citenamefont
  {Novoselov}, \citenamefont {Geim}, \citenamefont {Morozov}, \citenamefont
  {Jiang}, \citenamefont {Zhang}, \citenamefont {Dubonos}, \citenamefont
  {Grigorieva},\ and\ \citenamefont {Firsov}}]{Novoselov2004gate}%
  \BibitemOpen
  \bibfield  {author} {\bibinfo {author} {\bibfnamefont {K.~S.}\ \bibnamefont
  {Novoselov}}, \bibinfo {author} {\bibfnamefont {A.~K.}\ \bibnamefont {Geim}},
  \bibinfo {author} {\bibfnamefont {S.~V.}\ \bibnamefont {Morozov}}, \bibinfo
  {author} {\bibfnamefont {D.}~\bibnamefont {Jiang}}, \bibinfo {author}
  {\bibfnamefont {Y.}~\bibnamefont {Zhang}}, \bibinfo {author} {\bibfnamefont
  {S.~V.}\ \bibnamefont {Dubonos}}, \bibinfo {author} {\bibfnamefont {I.~V.}\
  \bibnamefont {Grigorieva}}, \ and\ \bibinfo {author} {\bibfnamefont {A.~A.}\
  \bibnamefont {Firsov}},\ }\href {\doibase 10.1126/science.1102896} {\bibfield
   {journal} {\bibinfo  {journal} {Science}\ }\textbf {\bibinfo {volume}
  {306}},\ \bibinfo {pages} {666} (\bibinfo {year} {2004})}\BibitemShut
  {NoStop}%
\bibitem [{\citenamefont {Wang}\ \emph {et~al.}(2008)\citenamefont {Wang},
  \citenamefont {Zhang}, \citenamefont {Tian}, \citenamefont {Girit},
  \citenamefont {Zettl}, \citenamefont {Crommie},\ and\ \citenamefont
  {Shen}}]{Wang2008gated-transistor}%
  \BibitemOpen
  \bibfield  {author} {\bibinfo {author} {\bibfnamefont {F.}~\bibnamefont
  {Wang}}, \bibinfo {author} {\bibfnamefont {Y.}~\bibnamefont {Zhang}},
  \bibinfo {author} {\bibfnamefont {C.}~\bibnamefont {Tian}}, \bibinfo {author}
  {\bibfnamefont {C.}~\bibnamefont {Girit}}, \bibinfo {author} {\bibfnamefont
  {A.}~\bibnamefont {Zettl}}, \bibinfo {author} {\bibfnamefont
  {M.}~\bibnamefont {Crommie}}, \ and\ \bibinfo {author} {\bibfnamefont
  {Y.~R.}\ \bibnamefont {Shen}},\ }\href {\doibase 10.1126/science.1152793}
  {\bibfield  {journal} {\bibinfo  {journal} {Science}\ }\textbf {\bibinfo
  {volume} {320}},\ \bibinfo {pages} {206} (\bibinfo {year}
  {2008})}\BibitemShut {NoStop}%
\bibitem [{\citenamefont {Chen}\ \emph {et~al.}(2012)\citenamefont {Chen},
  \citenamefont {Badioli}, \citenamefont {Alonso-Gonz{\'a}lez}, \citenamefont
  {Thongrattanasiri}, \citenamefont {Huth}, \citenamefont {Osmond},
  \citenamefont {Spasenovi{\'c}}, \citenamefont {Centeno}, \citenamefont
  {Pesquera}, \citenamefont {Godignon} \emph {et~al.}}]{Chen2012optical}%
  \BibitemOpen
  \bibfield  {author} {\bibinfo {author} {\bibfnamefont {J.}~\bibnamefont
  {Chen}}, \bibinfo {author} {\bibfnamefont {M.}~\bibnamefont {Badioli}},
  \bibinfo {author} {\bibfnamefont {P.}~\bibnamefont {Alonso-Gonz{\'a}lez}},
  \bibinfo {author} {\bibfnamefont {S.}~\bibnamefont {Thongrattanasiri}},
  \bibinfo {author} {\bibfnamefont {F.}~\bibnamefont {Huth}}, \bibinfo {author}
  {\bibfnamefont {J.}~\bibnamefont {Osmond}}, \bibinfo {author} {\bibfnamefont
  {M.}~\bibnamefont {Spasenovi{\'c}}}, \bibinfo {author} {\bibfnamefont
  {A.}~\bibnamefont {Centeno}}, \bibinfo {author} {\bibfnamefont
  {A.}~\bibnamefont {Pesquera}}, \bibinfo {author} {\bibfnamefont
  {P.}~\bibnamefont {Godignon}},  \emph {et~al.},\ }\href {\doibase
  10.1038/nature11254} {\bibfield  {journal} {\bibinfo  {journal} {Nature}\
  }\textbf {\bibinfo {volume} {487}},\ \bibinfo {pages} {77} (\bibinfo {year}
  {2012})}\BibitemShut {NoStop}%
\bibitem [{\citenamefont {Boukhvalov}\ and\ \citenamefont
  {Katsnelson}(2008)}]{Katsnelson2008defect}%
  \BibitemOpen
  \bibfield  {author} {\bibinfo {author} {\bibfnamefont {D.~W.}\ \bibnamefont
  {Boukhvalov}}\ and\ \bibinfo {author} {\bibfnamefont {M.~I.}\ \bibnamefont
  {Katsnelson}},\ }\href {\doibase 10.1021/nl802234n} {\bibfield  {journal}
  {\bibinfo  {journal} {Nano Letters}\ }\textbf {\bibinfo {volume} {8}},\
  \bibinfo {pages} {4373} (\bibinfo {year} {2008})}\BibitemShut {NoStop}%
\bibitem [{\citenamefont {Boukhvalov}\ and\ \citenamefont
  {Katsnelson}(2009)}]{Boukhvalov2009}%
  \BibitemOpen
  \bibfield  {author} {\bibinfo {author} {\bibfnamefont {D.~W.}\ \bibnamefont
  {Boukhvalov}}\ and\ \bibinfo {author} {\bibfnamefont {M.~I.}\ \bibnamefont
  {Katsnelson}},\ }\href {\doibase 10.1088/0953-8984/21/34/344205} {\bibfield
  {journal} {\bibinfo  {journal} {Journal of Physics: Condensed Matter}\
  }\textbf {\bibinfo {volume} {21}},\ \bibinfo {pages} {344205} (\bibinfo
  {year} {2009})}\BibitemShut {NoStop}%
\bibitem [{\citenamefont {Kuila}\ \emph {et~al.}(2012)\citenamefont {Kuila},
  \citenamefont {Bose}, \citenamefont {Mishra}, \citenamefont {Khanra},
  \citenamefont {Kim},\ and\ \citenamefont
  {Lee}}]{Kuila2012chemical-functionalization}%
  \BibitemOpen
  \bibfield  {author} {\bibinfo {author} {\bibfnamefont {T.}~\bibnamefont
  {Kuila}}, \bibinfo {author} {\bibfnamefont {S.}~\bibnamefont {Bose}},
  \bibinfo {author} {\bibfnamefont {A.~K.}\ \bibnamefont {Mishra}}, \bibinfo
  {author} {\bibfnamefont {P.}~\bibnamefont {Khanra}}, \bibinfo {author}
  {\bibfnamefont {N.~H.}\ \bibnamefont {Kim}}, \ and\ \bibinfo {author}
  {\bibfnamefont {J.~H.}\ \bibnamefont {Lee}},\ }\href {\doibase
  https://doi.org/10.1016/j.pmatsci.2012.03.002} {\bibfield  {journal}
  {\bibinfo  {journal} {Progress in Materials Science}\ }\textbf {\bibinfo
  {volume} {57}},\ \bibinfo {pages} {1061 } (\bibinfo {year}
  {2012})}\BibitemShut {NoStop}%
\bibitem [{\citenamefont {Vakil}\ and\ \citenamefont
  {Engheta}(2011)}]{Vakil2011}%
  \BibitemOpen
  \bibfield  {author} {\bibinfo {author} {\bibfnamefont {A.}~\bibnamefont
  {Vakil}}\ and\ \bibinfo {author} {\bibfnamefont {N.}~\bibnamefont
  {Engheta}},\ }\href {\doibase 10.1126/science.1202691} {\bibfield  {journal}
  {\bibinfo  {journal} {Science}\ }\textbf {\bibinfo {volume} {332}},\ \bibinfo
  {pages} {1291} (\bibinfo {year} {2011})}\BibitemShut {NoStop}%
\bibitem [{\citenamefont {Musa}\ \emph {et~al.}(2017)\citenamefont {Musa},
  \citenamefont {Renuka}, \citenamefont {Lin}, \citenamefont {Li},
  \citenamefont {Wang}, \citenamefont {Li}, \citenamefont {Zhang},\ and\
  \citenamefont {Chen}}]{Musa2017}%
  \BibitemOpen
  \bibfield  {author} {\bibinfo {author} {\bibfnamefont {M.~Y.}\ \bibnamefont
  {Musa}}, \bibinfo {author} {\bibfnamefont {M.}~\bibnamefont {Renuka}},
  \bibinfo {author} {\bibfnamefont {X.}~\bibnamefont {Lin}}, \bibinfo {author}
  {\bibfnamefont {R.}~\bibnamefont {Li}}, \bibinfo {author} {\bibfnamefont
  {H.}~\bibnamefont {Wang}}, \bibinfo {author} {\bibfnamefont {E.}~\bibnamefont
  {Li}}, \bibinfo {author} {\bibfnamefont {B.}~\bibnamefont {Zhang}}, \ and\
  \bibinfo {author} {\bibfnamefont {H.}~\bibnamefont {Chen}},\ }\href {\doibase
  10.1088/2053-1583/aa9643} {\bibfield  {journal} {\bibinfo  {journal} {2D
  Materials}\ }\textbf {\bibinfo {volume} {5}},\ \bibinfo {pages} {015018}
  (\bibinfo {year} {2017})}\BibitemShut {NoStop}%
\bibitem [{\citenamefont {Guti{\'{e}}rrez-Rubio}\ \emph
  {et~al.}(2013)\citenamefont {Guti{\'{e}}rrez-Rubio}, \citenamefont
  {Stauber},\ and\ \citenamefont {Guinea}}]{Guinea2013}%
  \BibitemOpen
  \bibfield  {author} {\bibinfo {author} {\bibfnamefont {A.}~\bibnamefont
  {Guti{\'{e}}rrez-Rubio}}, \bibinfo {author} {\bibfnamefont {T.}~\bibnamefont
  {Stauber}}, \ and\ \bibinfo {author} {\bibfnamefont {F.}~\bibnamefont
  {Guinea}},\ }\href {\doibase 10.1088/2040-8978/15/11/114005} {\bibfield
  {journal} {\bibinfo  {journal} {Journal of Optics}\ }\textbf {\bibinfo
  {volume} {15}},\ \bibinfo {pages} {114005} (\bibinfo {year}
  {2013})}\BibitemShut {NoStop}%
\bibitem [{\citenamefont {Bohm}\ and\ \citenamefont {Pines}(1951)}]{Pines1951}%
  \BibitemOpen
  \bibfield  {author} {\bibinfo {author} {\bibfnamefont {D.}~\bibnamefont
  {Bohm}}\ and\ \bibinfo {author} {\bibfnamefont {D.}~\bibnamefont {Pines}},\
  }\href {\doibase 10.1103/PhysRev.82.625} {\bibfield  {journal} {\bibinfo
  {journal} {Phys. Rev.}\ }\textbf {\bibinfo {volume} {82}},\ \bibinfo {pages}
  {625} (\bibinfo {year} {1951})}\BibitemShut {NoStop}%
\bibitem [{\citenamefont {Matsumoto}\ \emph {et~al.}(1980)\citenamefont
  {Matsumoto}, \citenamefont {Semenoff}, \citenamefont {Umezawa},\ and\
  \citenamefont {Tachiki}}]{Matsumoto1980}%
  \BibitemOpen
  \bibfield  {author} {\bibinfo {author} {\bibfnamefont {H.}~\bibnamefont
  {Matsumoto}}, \bibinfo {author} {\bibfnamefont {G.}~\bibnamefont {Semenoff}},
  \bibinfo {author} {\bibfnamefont {H.}~\bibnamefont {Umezawa}}, \ and\
  \bibinfo {author} {\bibfnamefont {M.}~\bibnamefont {Tachiki}},\ }\href
  {\doibase 10.1002/prop.19800280202} {\bibfield  {journal} {\bibinfo
  {journal} {Fortschritte der Physik}\ }\textbf {\bibinfo {volume} {28}},\
  \bibinfo {pages} {67} (\bibinfo {year} {1980})}\BibitemShut {NoStop}%
\end{thebibliography}%

\newpage 

\appendix
\begin{widetext}
\begin{center}
\textbf{Supplementary material}\\
Undamped transverse electric mode in undoped two-dimensional tilted Dirac cone materials\\
Z. Jalali-Mola and S. A. Jafari
\end{center}
\end{widetext}

In this supplementary material referred in the main text as SM, we review the textbook derivation
of the equations of TM and TE modes. This is true for any material. This SM is meant to make the
paper self-contained. Otherwise, the results are well-known and well established. We specialize
the results to 2D materials. 
\section{Transverse magnetic mode}{\label{A}}
In this section, we give more detail on the derivation of  transverse magnetic mode dispersion relation in TDMs. Supposing the 2D TDM located in $xy$ plane at $z=0$ and surrounded by two media with dielectric  constant $\ep_1, \ep_2$ and the propagation of electric and magnetic field along the $z$ axis, in surrounding media has evanescent behavior described with $\gamma$. 
As it has been mentioned in the body of paper electromagnetic TM mode propagates at interface with the magnetic field also lying in $xy$ plane and perpendicular to the 
to the wave vector. This implies $H_{i,z}=0$ in which $i=1,2$ stands for upper $z>0$ and lower half plane $z<0$.
\begin{align}
\label{TMBC.eqn}
&\bs H_i=(H_{i,x},H_{i,y},0)e^{i(\bsq.\bs r-\omega t)} e^{-\gamma_i |z|},\\
&\bs E_i=(E_{i,x},E_{i,y},E_{i,z})e^{i(\bsq.\bs r-\omega t)} e^{-\gamma_i |z|},\nn\\
&\bsq=(q_x,q_y)\nn
\end{align}
In the bulk of two media as a result of no free charge and current the Maxwell equation finds the following representation:
\bearr
&&\nabla \cdot \bs D=\rho \delta(z),~~~\nabla \cdot \bs B=0,\nn\\&&
\nabla \times \bs E=-\frac{1}{c}\pr_t \bs B,~~~
\nabla \times \bs B=\frac{1}{c}\pr_t \bs D.
\eearr
For ease of calculation we suppose two media are the same  without any magnetic properties hence we have, $\ep_1=\ep_2=\ep_0$ 
and $\mu_1=\mu_2=\mu_0$.  This properties imply that $\bs D_i=\ep_0 \bs E_i$ and $\bs H_i=\bs B_i/\mu_0$.
Hence the homogeneous equation for the electric and magnetic field in the bulk is given by,
\bearr
&&\nabla \times \nabla \times \bs E=-\frac{1}{c^2}\pr_t^2 \bs E.
\eearr
Inserting the definition of electric field components in above relation gives,

\be
\begin{bmatrix}
	q_y^2 -\gamma_i^2	&	 -q_x q_y	&	\mp iq_x \gamma \\
  	 -q_x q_y	&	q_x^2-\gamma_i^2	&		\mp iq_y \gamma\\
     \mp iq_x \gamma_i	&		  \mp iq_y \gamma_i		&	q_x^2+q_y^2\\
\end{bmatrix}
\bs E
-\frac{\omega^2}{c^{2}} \ep_i\bs E=0.
\ee

The determinant of above matrix for two media with the same dielectric constant defines the same decay constant  $\gamma_1^2=\gamma_2^2=q^2-\omega^2/c^2$. The dispersion of electromagnetic modes can be derived from boundary conditions at $z=0$ where the tilted Dirac plane is located.
\bearr
&&E_{1,t}=E_{2,t}\Rightarrow  E_{1,x}=E_{2,x} ~~ ,~~  E_{1,y}=E_{2,y},
\label{Et-bounary.eqn}\nn\\&&
D_{1,n}-E_{2,n}=4\pi \rho \delta(z) \Rightarrow E_{1,z}-E_{2,z}=4\pi \rho \delta(z),
\label{En-bounary.eqn}\nn\\&&
H_{1,n}=H_{2,n} \Rightarrow H_{1,z}=H_{2,z},
\label{Hn-boundary.eqn}\nn\\&&
H_{1,t}-H_{2,t}=\frac{4\pi}{c}\bs J \times \hat{n}  \Rightarrow
\nn\\&& H_{1,x}-H_{2,x}= \frac{4\pi}{c}J_y  ~~,~~H_{1,y}-H_{2,y}= -\frac{4\pi}{c}J_x.
\label{Ht-bounary.eqn}
\eearr

The current vector is related to the electric field vector by conductivity tensor as,
\be
\bs J=
\begin{bmatrix}
\sigma^{xx}	(\bsq ,\omega)&	 \sigma^{xy}(\bsq ,\omega)\\
  	\sigma^{yx}(\bsq ,\omega)	&  \sigma^{yy}(\bsq ,\omega)
\end{bmatrix}
\begin{bmatrix}
E_{x}	\\
 E_{y}	
\end{bmatrix}.
\ee
Moreover in each medium, the field equations are
\bearr
&&\nabla.\bs H=0 \Rightarrow iq_x H_{i,x}+iq_y H_{i,y}\mp \gamma_i H_{i,z}=0,\label{GaussH.eqn}\\&&
\nabla \times \bs E=-\frac{1}{c}\pr_t \bs B \Rightarrow 
i\omega H_{i,x}=\pm \gamma_i E_{i,y},
\nn\\&&
i\omega H_{i,y}=\pm \gamma_i E_{i,y},~ \omega H_{i,z}=q_x E_{i,y}+q_y E_{i,x},
\eearr
in which minus (plus) sign stands for $i=1(2)$.

Using  boundary condition  equation for normal component of electric field ($E_z$), Eq.~\eqref{Ht-bounary.eqn} and implementing with the continuity equation one finds,
\be
E_{1,z}-E_{2,z}=\frac{4\pi}{i \omega c^2} (q_xj_x+q_y j_y).
\label{TM-bound-Ez.eqn}
\ee
The Gauss's law for electric filed in each medium ($i=1,2$) can be rewritten as
\be
iq_x E_{i,x}+i q_y E_{i,y}\mp \gamma_i E_{i,z}=0.
\ee
Combination of Eqs.~\eqref{TM-bound-Ez.eqn} and~\eqref{TM-bound-Ez.eqn} gives rise to the following relation:
\bearr
&&E_{1,x}[q_x+\frac{2\pi i \gamma}{c^2 \omega}(q_x \sigma^{xx}+q_y \sigma^{yx})]\nn\\&&
+E_{1,y}[q_y+\frac{2\pi i \gamma}{c^2 \omega}(q_x \sigma^{xy}+q_y \sigma^{yy})]=0.
\label{boundaryEz.eqn}
\eearr
Using $\bs\nabla \cdot\bs B=0$ and Faraday's law we find,
\be
q_x E_{i,y}=q_y E_{i,x}.
\label{GaussH-TM.eqn}
\ee
Combination of \eqref{boundaryEz.eqn} and \eqref{GaussH-TM.eqn} gives the TM dispersion relation as
\be
1+\frac{2\pi i \gamma}{ \omega q^2}\Gamma(\bsq,\omega)=0,
\ee
where
\be
\Gamma(\bsq,\omega)= q_x^2\sigma^{xx}(\bsq,\omega)+q_y^2\sigma^{yy}(\bsq,\omega)+2 q_x q_y \sigma^{xy}(\bsq,\omega).
\ee
Inserting each element of conductivity tensor using $-i\omega \sigma^{ij}=e^2\Pi^{ij}$ and \eqref{xx-yy-xy-undoped.eqn} we find following dispersion for TM mode,
\be
1-\frac{2\pi e^2 \gamma}{ q^2} \Pi^{00}(\bs q,\Omega)=0
\ee
Not surprisingly, in the instantaneous limit $c\to \infty$ above TM dispersion relation reduces to that of plasmon in RPA.
\be
1-v(q) \Pi^{00}(\bs q,\Omega)=0
\ee
Indeed when the magnetic field is in the $xy$ ( 2D  matter ) plane the longitudinal component of electric field $E_{\parallel}$ which is parallel with propagation direction of electromagnetic field within dipole electric wave, induces longitudinal charge oscillation, well known as plasmon mode. 

\section{Transverse electric mode}{\label{B}}
 In the case of transverse electric field derivation, we supposed  the electric field is in the $xy$ surface ans is perpendicular to the propagation direction of electromagnetic wave, which implies $E_{i,z}=0$ in both media. Hence, generally the electric and magnetic field have the following representation as,
\bearr
\label{TEBC.eqn}
&&\bs E_i=(E_{i,x},E_{i,y},0)e^{i(\bsq.\bs r-\omega t)} e^{-\gamma_i |z|},\\&&
\bs H_i=(H_{i,x},H_{i,y},H_{i,z})e^{i(\bsq.\bs r-\omega t)} e^{-\gamma_i |z|},\nn\\&&
\bsq=(q_x,q_y)\nn,
\eearr
where $i=1,2$ stands for upper $z>0$ and lower half space $z<0$, as before, and $\gamma$ encodes the evanescent behavior of propagating electromagnetic modes along the $z$ direction.

The boundary condition for magnetic field in Eq.~\eqref{Hn-boundary.eqn}  and Faraday's law for normal component of magnetic field in each medium gives rise to the 
following relation,
\be
q_x H_{1,x}+q_y H_{1,y}=\frac{2\pi}{c} (q_x J_y-q_y J_x).
\ee
Inserting definition of $H_{1,x}$ and $H_{1,y}$
from Faraday's law, dispersion relation of TE mode can be derived as,
\be
1-\frac{2\pi i \omega}{c^2\gamma q^2}\Gamma'(\bsq,\omega),
\label{TE-dis-app.eqn}
\ee
in which
\be
\Gamma'(\bsq,\omega)=q_x^2 \sigma^{yy}(\bsq ,\omega)+ q_y^2 \sigma^{xx}(\bsq ,\omega)-2q_x q_y \sigma^{xy}(\bsq ,\omega).
\label{gamma-app.eqn}
\ee
It is also possible to represent above dispersion relation versus components of polarization tensor by using Eq.~\eqref{xx-yy-xy-undoped.eqn} of the main text.

\end{document}